\documentclass[a4paper,fleqn,usenatbib]{mnras}

\usepackage{newtxtext,newtxmath}

\usepackage[T1]{fontenc}
\usepackage{ae,aecompl}

\usepackage{graphicx}	
\usepackage{amsmath}	
\usepackage{amssymb}	
\usepackage{hyperref} 
\usepackage{tabularx} 

\title[Imprint of environment on stellar populations]{The weak imprint of environment on the stellar populations of galaxies}

\author[J. Trussler et al.]{James Trussler,$^{1,2}$\thanks{E-mail: jaat2@cam.ac.uk}
Roberto Maiolino,$^{1,2,3}$
Claudia Maraston,$^{4}$
Yingjie Peng,$^{5}$
Daniel Thomas,$^{4}$
\newauthor
Daniel Goddard$^{4}$
and Jianhui Lian$^{6}$
\\
$^{1}$Cavendish Laboratory, University of Cambridge, 19 J.J.\@ Thomson Avenue, Cambridge CB3 0HE, UK\\
$^{2}$Kavli Institute for Cosmology, University of Cambridge, Madingley Road, Cambridge CB3 0HA, UK\\
$^{3}$Department of Physics and Astronomy, University College London, Gower Street, London WC1E 6BT, UK\\
$^{4}$Institute of Cosmology and Gravitation, University of Portsmouth, Portsmouth PO1 3FX, UK\\
$^{5}$Kavli Institute for Astronomy and Astrophysics, Peking University, Beijing 100871, China\\
$^{6}$Department of Physics \& Astronomy, University of Utah, Salt Lake City, UT 84112, USA
}

\date{Accepted XXX. Received YYY; in original form ZZZ}

\pubyear{2020}

\begin{document}
\label{firstpage}
\pagerange{\pageref{firstpage}--\pageref{lastpage}}
\maketitle

\begin{abstract}
We investigate the environmental dependence of the stellar populations of galaxies in SDSS DR7.
Echoing earlier works, we find that satellites are both more metal-rich ($<0.1$~dex) and older ($<2$~Gyr) than centrals of
the same stellar mass. However, after separating star-forming, green valley and passive galaxies, we find
that the true environmental dependence of both stellar metallicity ($<0.03$~dex) and age ($<0.5$~Gyr) is in fact much weaker.
We show that the strong environmental effects found when galaxies are not differentiated result
from a combination of selection effects brought about by the environmental dependence of the quenched fraction of galaxies, and thus
we strongly advocate for the separation of star-forming, green valley and passive galaxies when the environmental dependence of galaxy properties are investigated.
We also study further environmental trends separately for both central and satellite galaxies.
We find that star-forming galaxies show no environmental effects, neither for centrals nor for satellites.
In contrast, the stellar metallicities of passive and green valley satellites increase weakly ($<0.05$~dex and $<0.08$~dex, respectively) with increasing halo mass, increasing local overdensity
and decreasing projected distance from their central; this effect is interpreted in terms of moderate
environmental starvation (`strangulation') contributing to the quenching of satellite galaxies.
Finally, we find a unique feature in the stellar mass--stellar metallicity relation for passive centrals, where galaxies in more massive haloes have larger
stellar mass ($\sim$$0.1$~dex) at constant stellar metallicity; this effect is interpreted in terms of dry merging of passive central galaxies and/or progenitor bias.
\end{abstract}

	


\section{Introduction}\label{sec:intro}

The environment in which galaxies form has long been thought to play an important role in shaping their evolution, as observations of galaxy clusters in the local Universe had revealed that early-type galaxies are preferentially found in dense environments, while late-type galaxies dominate the galaxy population in low density environments \citep{Dressler1980}. However, this prevalence of late-type galaxies in low-density environments could also be a mass-dependent effect, as late-type galaxies are preferentially low-mass \citep[e.g.\@][]{Wuyts2011}, and there are more low-mass galaxies in low density environments \citep[e.g.\@][]{Kauffmann2004, Yang2009}. Later studies \citep[e.g.\@][]{Baldry2006, Peng2010, Woo2013}, leveraging on the statistical power of large spectroscopic galaxy surveys, were able to disentangle the effects of mass and environment, confirming that the environment can play a significant role in galaxy quenching. Red, quiescent, early-type galaxies are indeed more likely to be found in dense regions, close to the centres of massive haloes in large galaxy clusters, while blue star-forming late-type galaxies tend to be found in underdense regions, in the outskirts of low-mass haloes. In particular, \citet{Peng2010} showed, for galaxies in the SDSS ($z < 0.1$) and zCOSMOS ($0.3 < z < 0.6$), that the effects of mass and environment are largely separable, implying that there are two distinct quenching processes at work: processes that primarily depend on mass (`mass quenching') and processes that primarily depend on environment (`environmental quenching'). \\
\indent Environmental quenching corresponds to external processes that quench star formation, through interactions between a galaxy and other galaxies, with the intracluster medium (ICM) and the gravitational potential of the dark matter halo of the group/cluster in which the galaxy is embedded. These physical processes preferentially operate in dense environments and therefore will be more important for galaxies residing in groups/clusters than for galaxies in the field. Satellite galaxies falling into a cluster can have the gas in their interstellar medium (ISM) rapidly removed as they move through the hot ICM, which can result in rapid quenching in a process known as ram pressure stripping \citep[e.g.\@][]{Gunn1972, Abadi1999}. Furthermore, strong, frequent tidal interactions between two close companion galaxies can also lead to the removal of gas \citep[`harassment', e.g.\@][]{Farouki1981, Moore1996}. Stripping of the circumgalactic medium (CGM), i.e.\@ the hot halo gas surrounding galaxies, will halt the accretion of cold gas on to the galaxy (known as starvation), therefore shutting down the supply of fuel for star formation, in a process known as `strangulation' \citep[e.g.\@][]{Larson1980, VanDenBosch2008}. Similarly, galaxies plunging in to a group/cluster and interacting with the group dark matter halo are likely to become detached from the cosmic filaments that feed galaxies with fresh gas from the IGM \citep[e.g.\@][]{Keres2005, Dekel2006, Dekel2009}, which, again, shuts down the fuel supply for star formation, in a process known as `cosmological starvation' \citep[e.g.\@][]{Feldmann2015a, VandeVoort2017b, AragonCalvo2019}.

Further insights into the impact of environment on galaxy evolution have been obtained by studying the chemical
enrichment of galaxies. Since the abundance of metals (elements heavier than helium) in the ISM results from
stellar nucleosynthesis, and is further affected by the ejection of metals by galactic winds and the accretion of
gas from the IGM, the metallicity of a galaxy traces both the star formation history and the flow of baryons into
and out of the galaxy. Observations of the gas-phase metallicities of galaxies in the local
Universe have revealed that satellite galaxies tend to be more metal-rich than central galaxies of the same stellar
mass \citep{Pasquali2012, Peng2014a} and that cluster galaxies tend to be more metal-rich than field galaxies
\citep{Ellison2009a}. Furthermore, at a fixed stellar mass, the gas-phase metallicity of satellite galaxies tends to increase with
increasing galaxy overdensity \citep[e.g.\@][]{Mouhcine2007, Cooper2008, Ellison2009a, Peng2014a, Wu2017a}, with
increasing halo mass \citep[e.g.\@][]{Pasquali2012} and with decreasing projected distance from the central galaxy
\citep[for massive clusters, e.g.\@][]{Petropoulou2012, Maier2016, Maier2019}. However, the dependence of gas-phase
metallicity with environment is typically rather weak, with most studies finding a $\sim$$0.05$~dex (at most
$\sim$$0.1$~dex) environmental effect \citep[see also][]{Hughes2013}.

In addition to gas-phase metallicities, the metallicity of the stars in galaxies can also be studied. Indeed, \citet{Pasquali2010} extensively investigated the dependence of the stellar metallicity of galaxies on the central--satellite dichotomy, as well as on group halo mass. They found that satellite galaxies tend to have higher stellar metallicities than central galaxies of the same stellar mass, with the stellar metallicity difference between centrals and satellites decreasing with increasing stellar mass. Furthermore, they also found that the stellar metallicities of satellite galaxies tend to increase with increasing halo mass, with the slope of the stellar mass--stellar metallicity relation becoming shallower with increasing halo mass.

A significant benefit of studying stellar metallicities is that they can also be reliably measured for passive galaxies \citep[e.g.\@][]{Thomas2005, Thomas2010, Gallazzi2014}. Thus, the chemical content in both actively star-forming and quenched galaxies can be compared. Such comparisons have not been possible for gas-phase metallicities, as the nebular emission in passive galaxies is often too weak (due to the lack of gas) and because proper calibration of gas-phase metallicity diagnostics in non star-forming regions have only recently become available \citep{Kumari2019}.

\citet{Peng2015} and \citet{Trussler2020a} studied the stellar metallicities of galaxies in the SDSS ($z\sim0$) and found that green valley and passive galaxies are significantly more metal-rich than star-forming galaxies of the same stellar mass, indicating that galaxies typically undergo significant chemical enrichment during the quenching phase. They argue that the stellar metallicity difference between star-forming and passive galaxies can be used to distinguish between different quenching mechanisms as the amount of chemical enrichment during the quenching phase depends on the mechanism. Using gas regulator models \citep{Peng2014b}, they find that the large stellar metallicity difference between star-forming and passive galaxies implies that for galaxies at all masses, quenching must have involved an extended phase of starvation.

In this paper we extend the analysis of \citet{Trussler2020a}, who studied the role of stellar mass in galaxy quenching by investigating the stellar metallicities and ages of galaxies. Here we instead shift our attention to the role of environment in galaxy quenching, by studying the imprint of the environment on the stellar populations of galaxies. 

This paper is structured as follows. In Section~\ref{sec:data}, we discuss the sample of galaxies, the galaxy parameters and the measures of environment that will be used in our analysis. In Section~\ref{sec:env_dependence}, we investigate the dependence of the stellar populations (i.e.\@ stellar metallicities and stellar ages) of galaxies on environment. In Section~\ref{sec:env_quenching}, we study how the stellar metallicity difference between star-forming and passive galaxies depends on environment, to investigate the role the environment plays in galaxy quenching. Finally, in Section~\ref{sec:conclusions}, we summarise our main findings and conclude. We assume that solar metallicity $\mathrm{Z}_\odot = 0.02$ throughout this work.

\section{Data}  \label{sec:data}

\subsection{Sample and galaxy parameters}

Here we give a brief description of the sample of galaxies and galaxy parameters used in our analysis. We refer the reader to \citet{Trussler2020a} for a more thorough description of the data. 

\subsubsection{Sample} \label{subsubsec:sample}

We use the spectroscopic sample of 930\,000 galaxies in the Sloan Digital Sky Survey Data Release 7 \citep[SDSS DR7,][]{York2000, Abazajian2009}, obtained using the Sloan 2.5~m telescope \citep{Gunn2006}. The spectra are in the optical/NIR (3800--9200 \AA), with a spectral resolution $R \sim 2000$ and a typical S/N $\sim 10$ for galaxies near the main sample flux limit. The SDSS sample suffers from incompleteness for $M_* < 10^{10}~\mathrm{M_\odot}$. In order to safely extend our analysis down to $M_* = 10^{9}~\mathrm{M_\odot}$, we apply the $V_{\mathrm{max}}$ weightings from \citet{Blanton2003} to correct for volume incompleteness. We restrict our analysis to galaxies with reliable spectroscopic redshifts in the range $0.02 < z < 0.085$ and with median S/N per spectral pixel > 20. The cut in redshift was applied to reduce the effect of cosmological evolution and aperture effects on the analysis, while the cut in S/N was applied to ensure that only galaxies with reliable stellar metallicities and stellar ages are studied in our analysis. As discussed in \citet{Trussler2020a}, such a high S/N cut could potentially introduce unwanted biases into our analysis, as low surface brightness galaxies are preferentially removed from the sample. While this should not be a concern for passive galaxies, it may potentially be problematic for low-mass ($M_* < 10^{10}~\mathrm{M_\odot}$) star-forming and green valley galaxies, for which only 7 per cent and 10 per cent of the sample have S/N $> 20$, respectively. However, similar to \citet{Trussler2020a}, we find that the choice of S/N cut adopted does not have a significant effect on our results ($<0.05$~dex on the stellar metallicities), with the key environmental trends discussed in this work holding true independent of the S/N criterion used.

\subsubsection{Galaxy parameters} \label{subsubsec:galaxy_params}

The spectral fitting code {\footnotesize FIREFLY} \citep{Comparat2017, Goddard2017b, Wilkinson2017} was used to obtain stellar metallicities and stellar ages for each galaxy in the SDSS sample. Briefly, {\footnotesize FIREFLY} is a $\chi ^2$ minimisation code that uses an arbitrarily weighted, linear combination of simple stellar populations (SSPs) to fit an input galaxy spectrum. The weighted sum of metallicities and ages of each of the SSPs is used to derive the stellar metallicity and stellar age of the galaxy. We make use of both mass-weighted and light-weighted stellar metallicities and ages in our analysis. The former are obtained by weighting each SSP by its stellar mass contribution, while the latter weights each SSP by its total luminosity across the fitted wavelength range (3500--7429 \AA). These two weightings are complementary as the mass-weighted quantities trace the cumulative evolution of the galaxy, while the light-weighted quantities primarily trace the properties of the younger (and therefore brighter) stellar populations. The SSPs used by {\footnotesize FIREFLY} were generated using the stellar population models of \citet{Maraston2011}, specifically their version based on the empirical stellar library MILES \citep{Sanchez-Blazquez2006} and using a Kroupa IMF \citep{Kroupa2001}. The \citet{Maraston2011}-MILES models have metallicities $[Z/\mathrm{H}] = -2.3,\, -1.3,\, -0.3,\, 0.0,\, 0.3$ and ages that span from 6.5~Myr to 15~Gyr, with a spectral resolution of 2.5~\AA \ FWHM and a wavelength coverage of 3500--7429~\AA. 

We use stellar masses $M_*$ and total star formation rates ($\mathrm{SFR}$s) from the publicly available MPA-JHU DR7 release of spectral measurements\footnote{The MPA-JHU data release is available at
\url{https://wwwmpa.mpa-garching.mpg.de/SDSS/DR7/}.}. The stellar
masses were obtained from fits to the photometry, using the Bayesian methodology of \citet{Kauffmann2003a}. In-fibre $\mathrm{SFR}$s were computed from the H$\alpha$ emission \citep{Brinchmann2004}, which were then aperture-corrected using photometry \citep{Salim2007} to obtain the total $\mathrm{SFR}$s.

We use the aforementioned stellar masses and total $\mathrm{SFR}$s, together with the bimodality in the $\mathrm{SFR}$--$M_*$ plane, to classify galaxies as star-forming, green valley and passive using the classifications from \citet{Trussler2020a}. Star-forming galaxies have relatively high specific star formation rates ($\mathrm{sSFR}$s), passive galaxies have relatively low $\mathrm{sSFR}$s, and green valley galaxies have $\mathrm{sSFR}$s that are intermediate between those of star-forming and passive galaxies. As a measure to exclude galaxies hosting an AGN (which may affect the determination of galaxy parameters), star-forming galaxies are further required to have their BPT classification set to `star-forming' according to the [N\thinspace {\scriptsize II}]-BPT diagram \citep{Brinchmann2004}. After applying our cuts on redshift, S/N and this selection criterion, our sample consists of 16\,685 star-forming galaxies, 8445 green valley galaxies and 53\,661 passive galaxies. 

\subsection{Environment measures} \label{subsubsec:env_measures}

In order to study the dependence of stellar metallicities and stellar ages on environment, we divide the galaxy population into centrals and satellites, using the galaxy group catalogue of \citet{Yang2005, Yang2007}.  Briefly, the catalogue is constructed using an iterative Friend-of-Friends algorithm that has been calibrated on mock catalogues \citep{Yang2005}. Galaxies that are sufficiently close in projected distance and redshift are initially assigned into tentative groups. The properties of the dark matter halo (e.g.\@ halo mass, virial radius, velocity dispersion) associated with each tentative group are determined and this information is then used to update group memberships. The halo properties are recomputed, and this procedure is repeated until there are no further changes to the group membership. There are three different group samples that are constructed from these galaxies. We make use of Sample I, which contains 599\,451 galaxies with SDSS redshifts only. In addition, for each sample, two group catalogues are constructed from the `Petrosian' and `Model' absolute magnitudes of the galaxies in the NYU Value-Added Galaxy Catalog \citep[NYU-VAGC,][]{Blanton2005a}, respectively. We use the `Model' magnitudes in our analysis. We find that using the other sample and magnitude combinations has no significant effect on our results. 

\begin{figure*}
\centering
\centerline{\includegraphics[width=\linewidth]{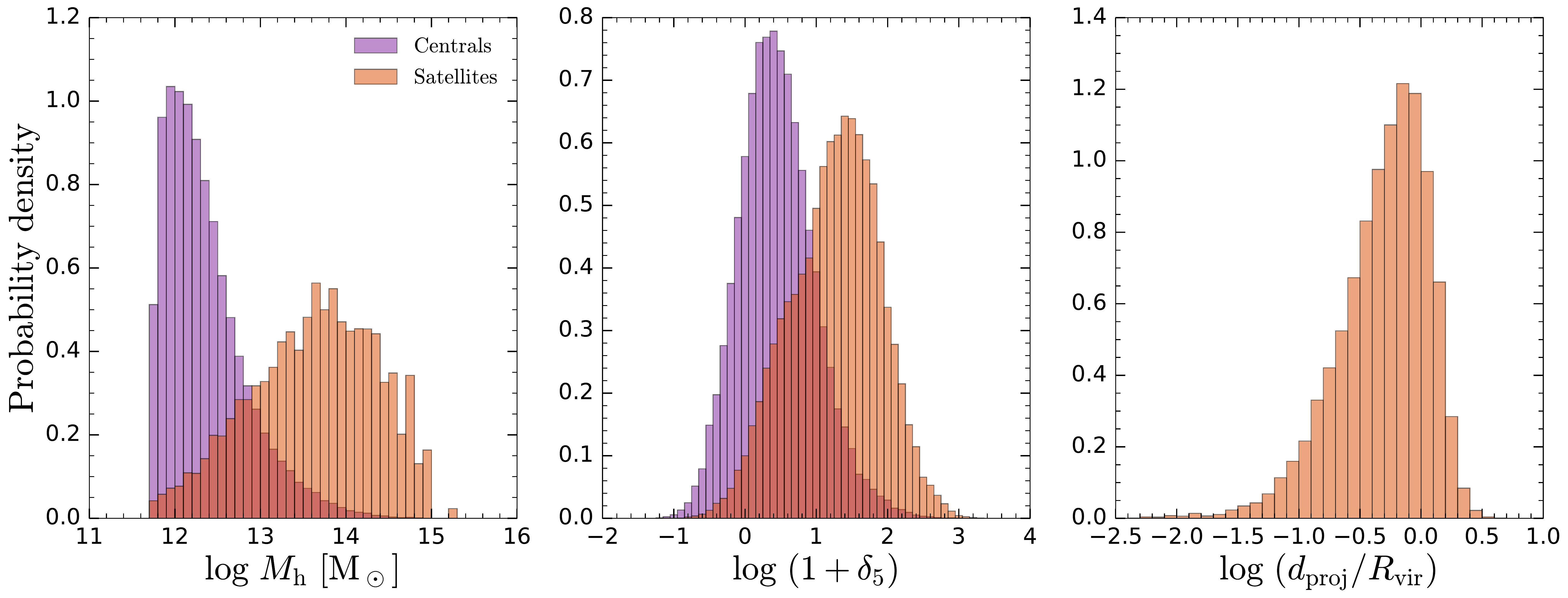}}
\caption{Distribution functions for all the centrals (purple) and satellites (orange) in our sample, in terms of group halo mass $M_\mathrm{h}$ (left panel), local overdensity $\delta_5$ (middle panel) and projected distance $d_\mathrm{proj}/R_\mathrm{vir}$ (only for satellites, right panel). The projected distances for centrals are not shown since they are, by definition, at the centres of their respective groups.}
\label{fig:cent_sat_distributions}
\end{figure*}

Having established groups and group membership, we then determine whether galaxies are centrals or satellites. Central galaxies are defined to be the most massive in their group. All other members of the group are defined to be satellite galaxies. We do not take the galaxy's spatial position in the group into account in our central--satellite classification system. Under the central--satellite definitions that we have used, isolated galaxies are always classified as centrals. Since there are a large number of groups with only a single galaxy (421\,067 out of 472\,504) in the \citet{Yang2007} group catalogue, a significant fraction  of the central galaxies in our study are isolated centrals.

We also make use of the \citet{Yang2007} group halo masses $M_\mathrm{h}$. These halo masses were estimated using abundance matching applied in two different ways. First, by ranking the galaxy groups according to their total luminosity. Second, by ranking the galaxy groups according to their total stellar mass. Since stellar mass estimates are less strongly affected by recent star formation activity than luminosity, we have chosen to use the stellar-mass ranked halo masses in our analysis. We note that for some small groups (in terms of group size, and stellar mass of the central galaxy), no halo masses are assigned. We exclude these groups from our halo mass analysis. This cut preferentially removes low-mass centrals (since these tend to be in small groups) from the sample, but only has a relatively minor effect on the satellite subsample. 

We also investigate how the properties of satellites are affected by their distance from the group centre, which we define to be the position of the central galaxy in the group. For each satellite galaxy, the projected distance $d_\mathrm{proj}$ from the central in its group is computed. This distance is then normalised by the virial radius $R_\mathrm{vir}$ of the group, which is given by $R_\mathrm{vir} = 120(M_\mathrm{h}/10^{11}~M_\odot)^{1/3}$~kpc \citep{Dekel2006}. We exclude satellites that reside in groups with no assigned halo mass from our projected distance analysis, as it is not possible to determine the virial radius of the group in this case.

To further investigate the role environment plays in galaxy evolution, we study the local overdensity, which is a dimensionless density contrast given by $\delta _i  = (\rho_i - \rho_\mathrm{m})/\rho_\mathrm{m}$, where $\delta_i$ is the overdensity around the $i^{\mathrm{th}}$ galaxy, $\rho_i$ is an estimate of the local density around the $i^{\mathrm{th}}$ galaxy and $\rho_\mathrm{m}$  is the mean density at that redshift. We use the overdensities estimated by \citet{Peng2010}, which were computed following the methodology developed by \citet{Kovac2010}. For each galaxy, the projected distance to the fifth nearest neighbour $d_5$ that lies within $\pm$1000~km s$^{-1}$ of that galaxy is determined. The local density $\rho_i$ is then calculated by dividing the number of neighbouring galaxies (5) by the volume of a cylinder of radius $d_5$ and with length (in the radial direction) $\pm$1000 km s$^{-1}$. \\
\indent After matching the SDSS DR7 subsample discussed in Section~\ref{subsubsec:galaxy_params} with both the \citet{Yang2007} catalogue and the overdensity dataset, our sample consists of 43\,538 centrals and 16\,805 satellites. For our halo mass and projected distance analysis, the sample size is further reduced to 35\,440 centrals and 16\,434 satellites, as groups that do not have halo masses assigned are excluded.\\
\indent We show the distribution functions for all the centrals (purple) and satellites (orange) in our sample, in terms of group halo mass $M_\mathrm{h}$ (left panel), local overdensity $1 + \delta _5$ (middle panel) and projected distance $d_\mathrm{proj}/R_\mathrm{vir}$  (only for satellites, right panel) in Fig.\@~\ref{fig:cent_sat_distributions}. The projected distances for centrals are not shown since they are, by definition, at the centres of their respective groups. We find that satellites tend to be, on average, in more massive haloes and denser regions than centrals. The same trends hold when comparing centrals and satellites in a narrow range of stellar mass (not shown). These trends come about because centrals of a given stellar mass occupy a narrow range in halo mass \citep[see e.g.\@][]{Yang2008, Yang2009, Yang2012}. By our central--satellite definition, satellites with that stellar mass will have to have an even more massive central, and hence will occupy a group with a higher halo mass. Since the local overdensity tends to increase with halo mass \citep[e.g.\@][]{Woo2013}, satellites will therefore also tend to be in denser regions than centrals of the same stellar mass. 

\section{The dependence of stellar populations on environment} \label{sec:env_dependence}

In this section we investigate how the stellar populations of galaxies, i.e.\@ stellar metallicities and stellar ages, and their relations to stellar mass, depend on environment. We will explore the similarities and differences between centrals and satellites in Section~\ref{subsec:cent_sat_dichotomy}, the dependence of the stellar mass--stellar metallicity relation on halo mass in Section~\ref{subsec:sp_hm}, the dependence on the local overdensity in Section~\ref{subsec:sp_od} and the dependence on projected distance in Section~\ref{subsec:sp_proj}.

\subsection{Central--satellite dichotomy} \label{subsec:cent_sat_dichotomy}

In this section, we study the dependence of the stellar mass--stellar metallicity relation and the stellar mass--stellar age relation on the central--satellite dichotomy. Galaxies are classified as either central or satellite using the definitions in Section~\ref{subsubsec:env_measures}, are subsequently binned in stellar mass, and both the median stellar metallicity and the error on the median ($1.253\sigma /\sqrt{N}$) are computed for each mass bin, where $\sigma$ is the standard deviation of the stellar metallicity in the mass bin, and $N$ is the number of galaxies in that mass bin. 

\subsubsection{Stellar metallicity} \label{subsubsec:cent_sat_mzr}

\begin{figure*}
\centering
\centerline{\includegraphics[width=1\linewidth]{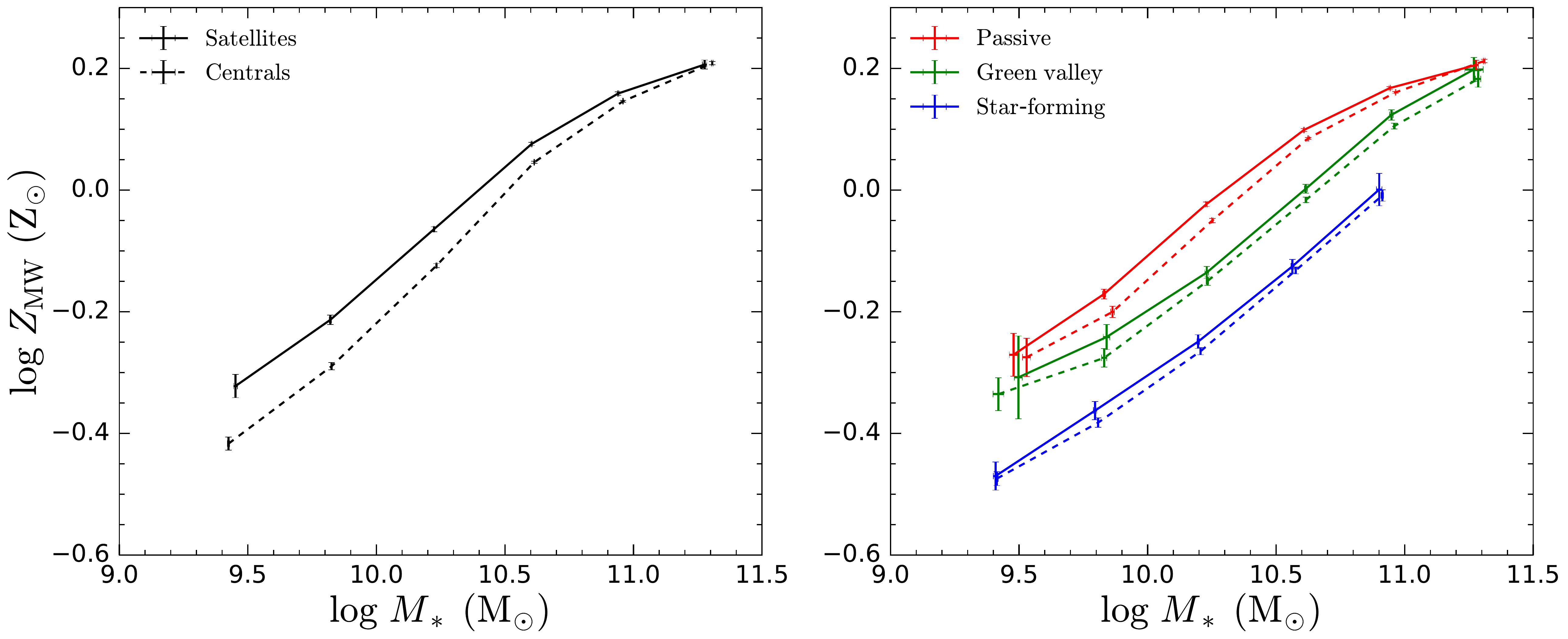}}
\caption{The mass-weighted stellar mass--stellar metallicity relation for centrals (dashed) and satellites (solid). Galaxies are binned in stellar mass and the median stellar metallicity in each mass bin is shown. Error bars correspond to the $1\sigma$ uncertainty on the median. Left panel: Including all galaxies (i.e.\@ without splitting into star-forming, green valley and passive galaxies, black). Right panel: Separating into star-forming (blue), green valley (green) and passive galaxies (red).}
\label{fig:mzr_mw_cent_sat}
\end{figure*}

We show the mass-weighted stellar mass--stellar metallicity relation for centrals (dashed) and satellites (solid) in the left panel of Fig.\@~\ref{fig:mzr_mw_cent_sat}. We find that satellites are typically more metal-rich than centrals of the same stellar mass, and that this metallicity difference is highly significant (i.e.\@ multiple $\sigma$). Furthermore, the metallicity difference between centrals and satellites decreases with increasing stellar mass, from $\sim$$0.1$~dex at the low-mass end down to $\sim$$0$ at the high-mass end. Taken at face value, this large offset between centrals and satellites suggests that the gas flow and chemical enrichment histories of central and satellite galaxies may be different, with the environment playing a relatively more important role at shaping galaxy evolution at the low-mass end.

Our result is qualitatively consistent with \citet{Pasquali2010} \citep[who studied the smaller sample of galaxies in SDSS DR4,][]{Adelman-McCarthy2006}, although their results were based on light-weighted stellar metallicities (rather than
the mass-weighted stellar metallicities used in our work), that were obtained using a different spectral fitting
procedure (through simultaneous fitting of five metallicity- and age-sensitive optical spectral absorption
features from \citet{Gallazzi2005}, rather than through a full spectral fit of the optical spectrum using {\footnotesize
FIREFLY}). We study the stellar mass--stellar metallicity relation for centrals and satellites using light-weighted metallicities from {\footnotesize FIREFLY} and from \citet{Gallazzi2005} in Appendix~\ref{sec:mzr_lw}. Briefly, we find similar qualitative trends to the results obtained using mass-weighted stellar metallicities, although the quantitative details are slightly different. 

However, as previously shown in \citet{Peng2015} and \citet{Trussler2020a}, and also depicted in the right panel of Fig.\@~\ref{fig:mzr_mw_cent_sat}, passive galaxies (shown in red) and green valley galaxies (shown in green) are substantially more metal-rich than star-forming galaxies (shown in blue) of the same stellar mass. Hence, if the relative abundance of star-forming, green valley and passive galaxies (i.e.\@ the quenched fraction, or non-SF fraction) is different for centrals and satellites, then one would expect the average stellar metallicity for the overall populations (i.e. including star-forming, green valley and passive galaxies together) of centrals and satellites to be different, because they have different proportions of relatively metal-poor star-forming and relatively metal-rich green valley/passive galaxies. So, even if the stellar metallicity of star-forming centrals and satellites, green valley centrals and satellites, and passive centrals and satellites are the same, one would still expect to see a metallicity offset between the overall populations of centrals and satellites because of this effect, which we shall refer to as the quenched fraction effect. Thus, the large offset between the overall population of centrals and satellites shown in the left panel of Fig.\@~\ref{fig:mzr_mw_cent_sat} may in fact be driven by a strong dependence of quenched fraction on environment, rather than a strong dependence of stellar metallicity on environment.

\begin{figure}
\centering
\centerline{\includegraphics[width=\linewidth]{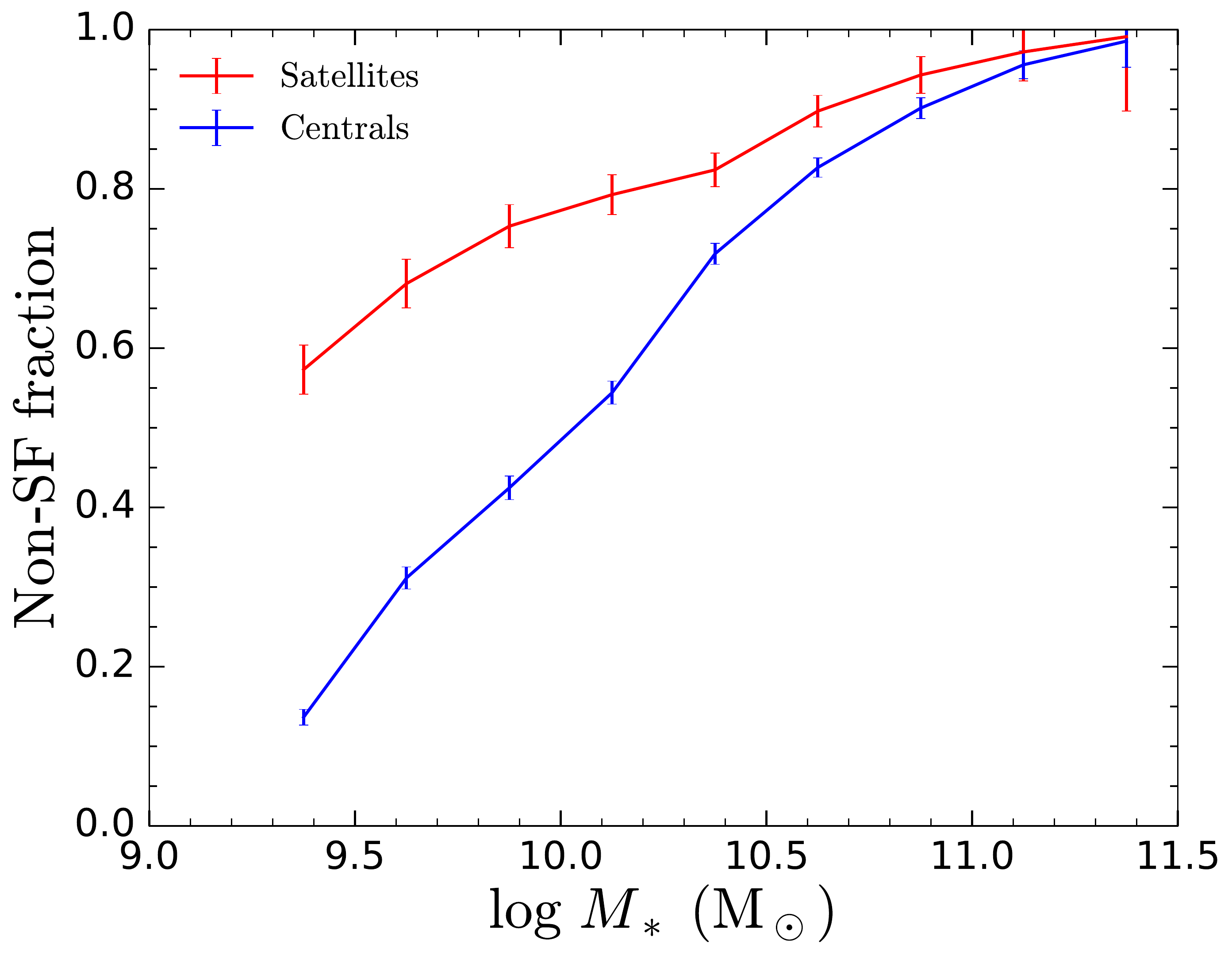}}
\caption{The non-SF fraction (i.e.\@ the fraction of galaxies that are green valley or passive) for centrals (blue) and satellites (red) in our sample (which has been biased by cuts on S/N). The errors on the non-SF fractions were obtained by propagating the errors (assumed to be Poissonian) on the green valley + passive galaxy count and the total galaxy count in each mass bin.
}
\label{fig:non_sf_fractions}
\end{figure}

The dependence of quenched fraction on the environment has been extensively investigated in the literature
\citep[e.g.\@][]{Peng2010, Peng2012, Woo2013, Wang2018a}. In particular, it has been found that, at a fixed stellar
mass, the quenched fraction for satellites is higher than for centrals. For the sake of convenience, we show this
result using our sample (which has been biased by cuts on S/N), but showing the non-SF fraction (i.e.\@ including green valley and passive galaxies), rather than the quenched fraction, in Fig.\@~\ref{fig:non_sf_fractions}. Since the non-SF fraction is higher for satellites than for centrals of the same stellar mass, the satellite population consists of relatively more green valley and passive galaxies, which have high stellar metallicities, and relatively fewer star-forming galaxies, which have lower stellar metallicities. Thus, at a given stellar mass, one would expect the overall satellite population to have a higher stellar metallicity than the overall central population, which is what was seen in the left panel of Fig.\@~\ref{fig:mzr_mw_cent_sat}.

In order to disentangle the inherent dependence of stellar metallicity on environment from the trends brought about by the dependence of quenched fraction on environment, we further divide the central and satellite populations into star-forming, green valley and passive galaxies in the right panel of Fig.\@~\ref{fig:mzr_mw_cent_sat}. We find that the stellar metallicity difference between centrals and satellites that was seen when considering the overall population is greatly reduced after separating into star-forming, green valley and passive galaxies. Evidently the dependence of quenched fraction on environment exaggerates the true stellar metallicity difference between central and satellite galaxies, which is actually much smaller (and of less significance), with satellites only being marginally more metal-rich than centrals of the same stellar mass.

\begin{figure}
\centering
\centerline{\includegraphics[width=\linewidth]{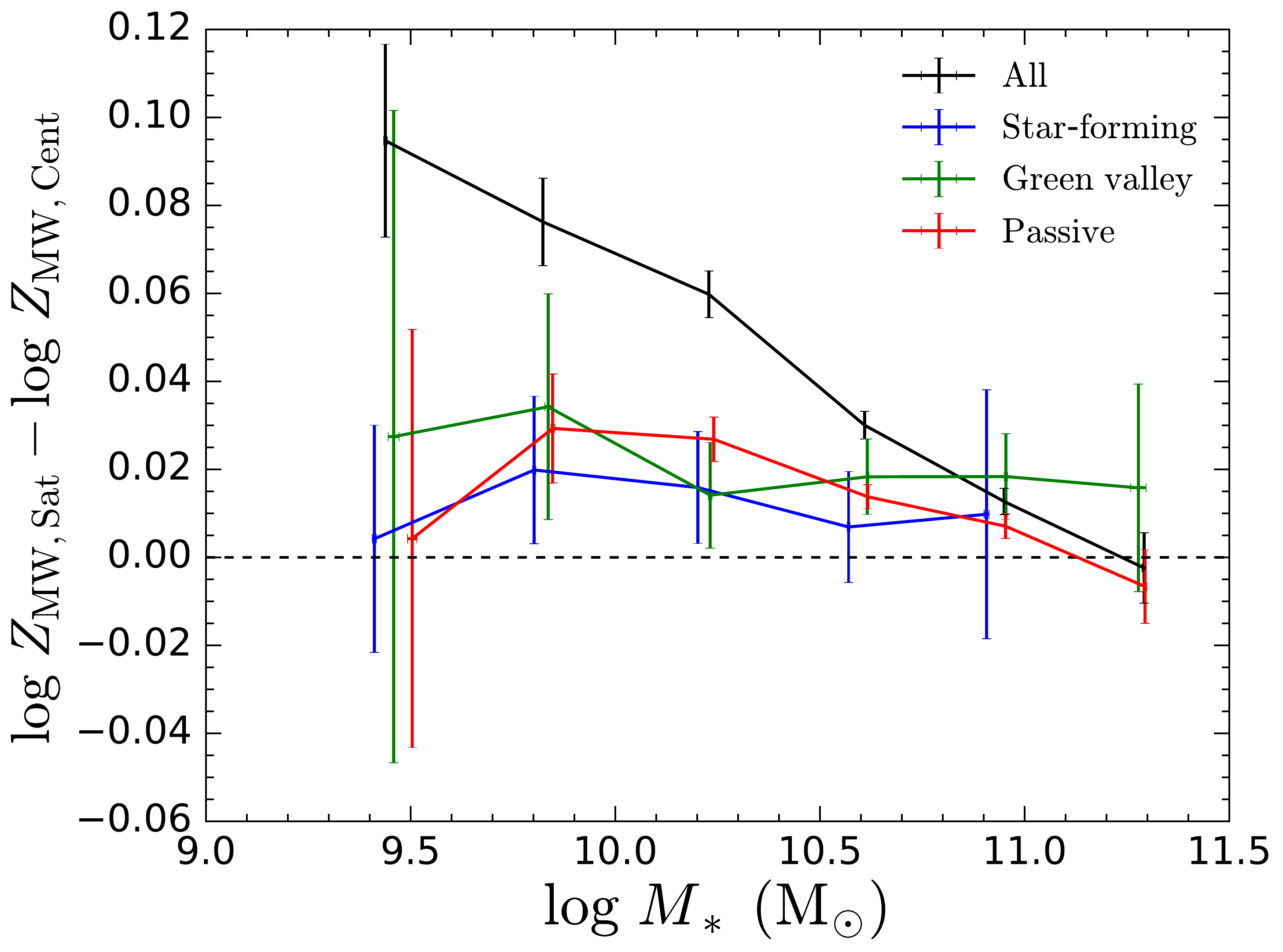}}
\caption{The mass-weighted stellar metallicity difference $\Delta \log Z_\mathrm{MW} = \log Z_\mathrm{MW, Sat} - \log Z_\mathrm{MW, Cent}$ between centrals and satellites of the same stellar mass, including all galaxies (i.e.\@ without splitting into star-forming, green valley and passive galaxies, black), for star-forming galaxies (blue), green valley galaxies (green) and passive galaxies (red). Error bars correspond to the $1\sigma$ error on the stellar metallicity difference.}
\label{fig:delta_mzr_cents_sats}
\end{figure}

We show the stellar metallicity difference between centrals and satellites more clearly in Fig.\@~\ref{fig:delta_mzr_cents_sats}. The typical metallicity difference for star-forming galaxies is $0.01$--$0.02$~dex, for green valley galaxies is $0.02$--$0.03$~dex and for passive galaxies is $0$--$0.03$~dex, which are clearly much smaller than what is seen when simply comparing the overall central population against the overall satellite population (shown in black), except perhaps at the high-mass end ($\log \, (M_*/\mathrm{M}_\odot) \geq 11.0$). 

Our results for star-forming, green valley and passive galaxies are qualitatively consistent with the findings of \citet{Bahe2017}, who used the cosmological hydrodynamical EAGLE simulation to study the stellar metallicities of $>3600$ galaxies with $\log \, (M_*/\mathrm{M}_\odot) > 10$. Similar to this work, they find that star-forming satellites, green valley satellites and passive satellites are more metal-rich than star-forming centrals, green valley centrals and passive centrals of the same stellar mass, respectively. \citet{Bahe2017} argue that the enhanced stellar metallicity of satellite galaxies is due to the fact that the star-forming gas in satellites is relatively more metal-rich than for centrals of the same stellar mass, which in turn is driven by (ram pressure) stripping of low-metallicity gas from the outskirts of satellites, together with the suppression of metal-poor inflows towards the centres of satellites.

However, there are quantitative differences with our work, as they typically find larger offsets between centrals and satellites, with a typical metallicity difference of $0.01$--$0.03$~dex for star-forming galaxies, $0.05$--$0.07$~dex for green valley galaxies and $0.01$--$0.08$~dex for passive galaxies. In contrast to our results, they find that the metallicity difference between the overall population of centrals and satellites is intermediate between (rather than greater than) the metallicity differences for star-forming, green valley and passive galaxies, at $0.02$--$0.05$~dex. This disagreement with our work is likely to be related to the quenched fraction effect, i.e.\@ the selection effect brought about by the difference in quenched fraction between centrals and satellites, that occurs when the overall populations of centrals and satellites are compared, which exaggerates and/or misrepresents the true difference in the properties of centrals and satellites, due to the starkly different properties (e.g.\@ stellar metallicities) of star-forming and passive galaxies \citep[see e.g.\@][]{Peng2015, Trussler2020a}. If, in the EAGLE simulation, either the stellar metallicities of star-forming, green valley and passive galaxies are very similar, or the quenched fractions of centrals and satellites are similar, then the quenched fraction effect will be minimal and one should obtain the \citet{Bahe2017} result for the overall population of centrals and satellites. Indeed, using $\mathrm{sSFR}$s to classify EAGLE galaxies as star-forming, green valley and passive, we find that the stellar metallicity differences $\Delta \log Z_*$ between star-forming and passive galaxies in the EAGLE simulation ($< 0.03$~dex) are in fact much smaller than what was found observationally using SDSS data ($\sim$$0.2$~dex). Thus, while the quenched fraction effect greatly exaggerates the true differences between centrals and satellites in the observational data (due to the large $\Delta \log Z_*$ between star-forming and passive galaxies), it only has a minimal effect in the EAGLE data (due to the small $\Delta \log Z_*$).

Furthermore, the quenched fraction effect also misrepresents the mass-dependence of the true metallicity difference between centrals and satellites. While the metallicity difference for the overall population clearly declines with increasing stellar mass, it is roughly independent of stellar mass for star-forming and green valley galaxies, and appears to decline more weakly with increasing stellar mass for passive galaxies. Thus, the strong mass dependence of the stellar metallicity difference for the overall population (shown in Figs.\@ \ref{fig:mzr_mw_cent_sat} and \ref{fig:delta_mzr_cents_sats}) is not driven by the inherent dependence of stellar metallicity on environment (i.e.\@ centrals vs satellites), but rather on the relative mass dependence of the non-SF fractions for centrals and satellites (shown in Fig.\@~\ref{fig:non_sf_fractions}).

To summarise this section, the stellar metallicity difference between the overall population of centrals and satellites is
mostly brought about by their difference in quenched fraction. Owing to the large offset in stellar metallicity between star-forming, green valley and passive galaxies, the quenched fraction effect is much more prominent than the inherent environmental dependence in driving the apparent metallicity--environment relations. We therefore strongly suggest that the quenched fraction effect is taken into account (by splitting into star-forming,
green valley and passive galaxies) whenever environmental trends are investigated, especially for scaling relations that are different for star-forming and passive galaxies.

\subsubsection{Stellar age}

\begin{figure*}
\centering
\centerline{\includegraphics[width=1\linewidth]{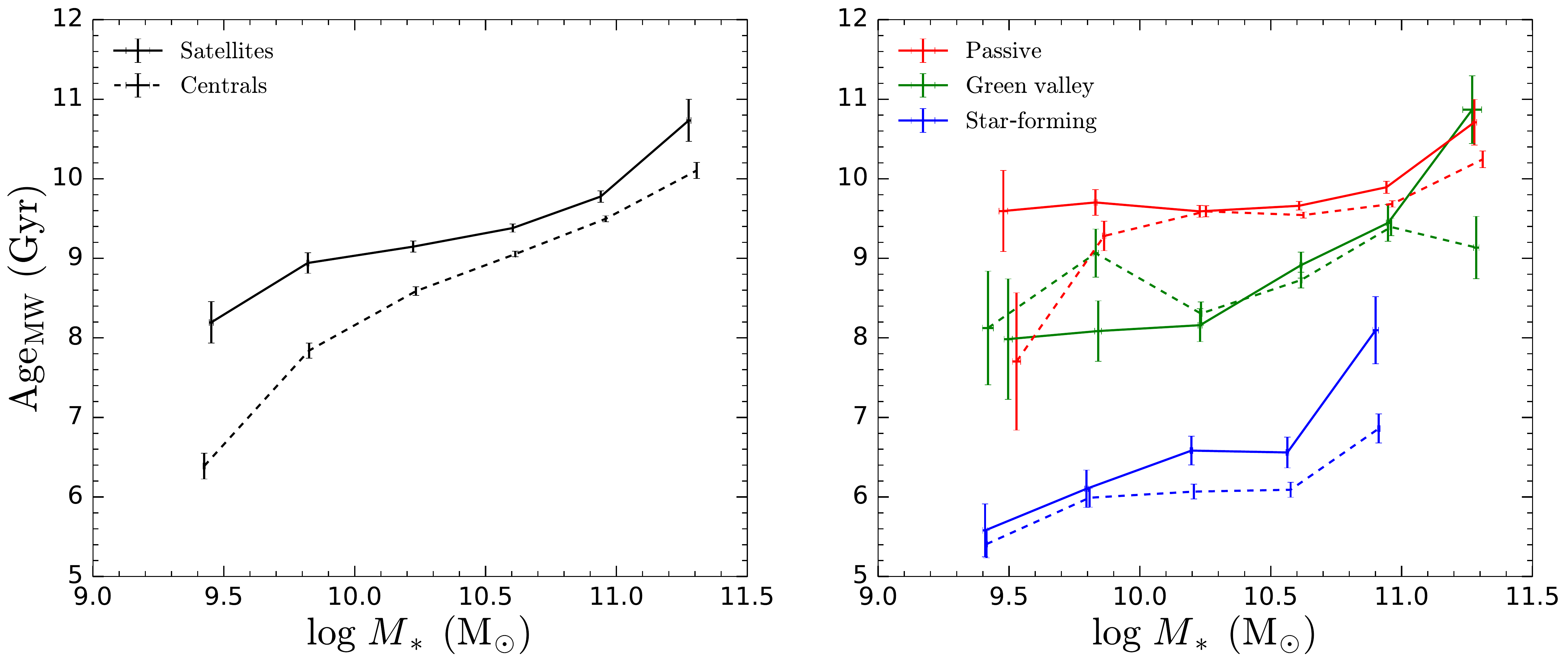}}
\caption{Similar to Fig.\@~\ref{fig:mzr_mw_cent_sat}, but now showing the mass-weighted stellar mass--stellar age relation.}
\label{fig:mar_mw_cent_sat}
\end{figure*}

We show the mass-weighted stellar mass--stellar age relation for the overall central population (dashed) and
satellite population (solid) in the left panel of Fig.\@~\ref{fig:mar_mw_cent_sat}. We find that satellites are
typically older than centrals of the same stellar mass, with the difference being highly significant (i.e.\@ multiple $\sigma$) across the entire stellar mass range. In addition, the stellar age difference decreases with increasing stellar mass, from $\sim$2~Gyr at $\log \, (M_*/\mathrm{M}_\odot) = 9.5$ to $\sim$0.5~Gyr at $\log \, (M_*/\mathrm{M}_\odot) = 11.5$. Again, we find that our results are qualitatively consistent with the work of \citet{Pasquali2010}, who obtained similar trends with SDSS DR4, using light-weighted ages and a different spectral fitting procedure \citep[see][]{Gallazzi2005}. We briefly study the light-weighted stellar mass--stellar age relation using light-weighted ages from {\footnotesize FIREFLY} and \citet{Gallazzi2005} in Appendix~\ref{sec:mar_lw}, and find that the qualitative (but not the quantitative) trends obtained in this section are preserved. Taken at face value, the large age offset between centrals and satellites suggests that the star formation histories of central and satellite galaxies may be different, with the imprint of environment being strongest at the low-mass end and weaker at the high-mass end.

However, as shown in the right panel of Fig.\@~\ref{fig:mar_mw_cent_sat}, after splitting into star-forming, green valley and passive galaxies, the age differences between centrals and satellites tend to become much smaller and less significant than what was seen for the overall population. In some cases the age differences are still large, such as the lowest mass bin for passive galaxies and the highest mass bin for green valley galaxies, but it should be noted that the statistics in these mass bins are very low, so the large offset may partly be due to noise. Similar to what was seen for the overall population, star-forming satellites tend to be older than star-forming centrals, and passive satellites tend to be older than passive centrals of the same stellar mass. The trend for green valley galaxies is less clear, which may be because the statistics for green valley galaxies are relatively low with respect to star-forming and passive galaxies (see our sample description in Section~\ref{subsubsec:galaxy_params}). Neglecting the highest mass bin, we find that at high masses the ages of green valley satellites and green valley centrals are comparable.\footnote{On the other hand, at the low-mass end there is even an indication that green valley centrals are older than green valley satellites, which is a reversal of the trend seen elsewhere (satellites are older than centrals). However, we do not find this result to be significant, since the stellar mass--stellar age relation is, roughly speaking, a monotonically increasing relation, so the relatively high age for centrals in the second-lowest mass bin may be due to random scatter driven by the low-number statistics.}

Furthermore, while the age difference between the overall population of centrals and satellites clearly decreases with increasing stellar mass, there are no clear trends for star-forming, green valley and passive galaxies. If anything, star-forming galaxies exhibit the opposite trend, with the age difference between centrals and satellites instead increasing with increasing stellar mass.

\subsection{Halo mass} \label{subsec:sp_hm}

Having studied the dependence of the mass-weighted stellar mass--stellar metallicity relation on the central--satellite dichotomy, we now move on to further measures of the environment. In this section we study how the mass-weighted stellar mass--stellar metallicity relation depends on group halo mass $M_\mathrm{h}$ for both centrals and satellites. 

We bin galaxies into quartiles of halo mass, which we refer to as Low, Mid-Low, Mid-High and High, for the 1st (i.e.\@ the least massive haloes), 2nd, 3rd and 4th (i.e.\@ the most massive haloes) quartiles, respectively. Since the distributions of centrals and satellites in halo mass are different (see Fig.\@~\ref{fig:cent_sat_distributions}), we use different halo mass quartiles for centrals and satellites to more clearly identify the trends with halo mass. To clarify, this means that the halo mass range associated with e.g.\@ the Low quartile is different for centrals and satellites. Furthermore, since passive galaxies are more likely to reside in massive haloes, while star-forming galaxies are more likely to reside in low-mass haloes \citep{Woo2013}, the distributions for star-forming, green valley and passive galaxies in halo mass are different. Hence we also use different halo mass quartiles for star-forming galaxies, green valley galaxies, passive galaxies and the overall population of galaxies.

\subsubsection{Satellites}

\begin{figure*}
\centering
\centerline{\includegraphics[width=1\linewidth]{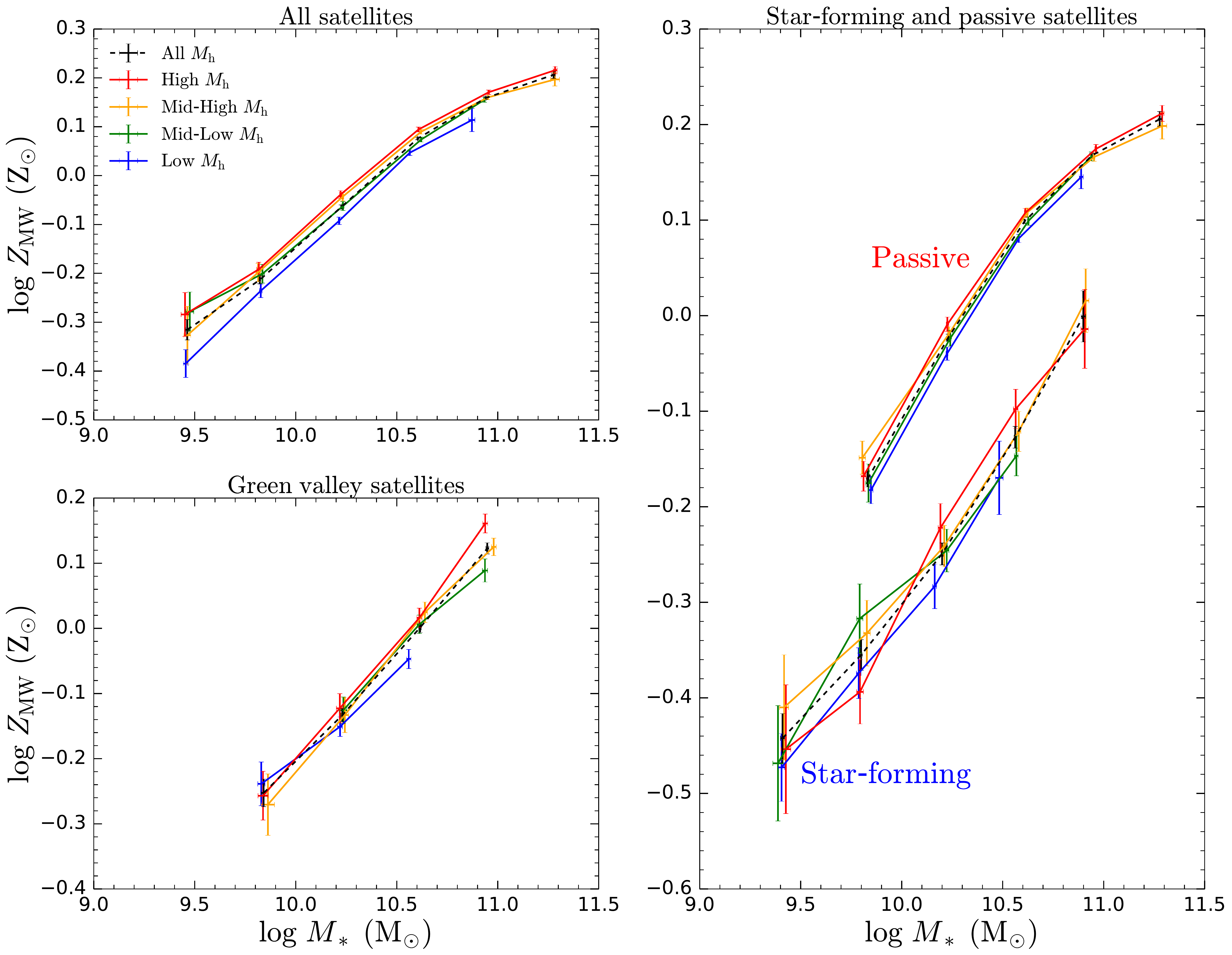}}
\caption{The mass-weighted stellar mass--stellar metallicity relation for satellites in different quartiles of halo mass $M_\mathrm{h}$, which we refer to as Low (shown in blue), Mid-Low (green), Mid-High (orange) and High (red) for the 1st (i.e.\@ the least massive haloes), 2nd, 3rd and 4th (i.e.\@ the most massive haloes) quartiles, respectively, as well as the relation obtained without splitting into quartiles of halo mass (named All, in dashed black). Note that the vertical scale and range is different for each panel. Furthermore, the halo mass quartiles in the different panels correspond to different halo mass ranges. Top-left panel: all satellites (i.e.\@ without splitting into star-forming, green valley and passive galaxies). Right panel: star-forming and passive satellites. Bottom-left panel: green valley satellites.}
\label{fig:halo_mzr_mw_sat}
\end{figure*}

We show the mass-weighted stellar mass--stellar metallicity relation for the overall population of satellites in different halo mass quartiles in the top-left panel of Fig.\@~\ref{fig:halo_mzr_mw_sat}. We find that, at a fixed stellar mass, the stellar metallicity for the overall population of satellites tends to increase with increasing halo mass, and that this applies across the entire stellar mass range studied. Although the stellar metallicity difference between adjacent quartiles is small and perhaps marginally significant, there is a clear offset between the lowest and highest halo mass quartiles, that tends to decrease with increasing halo mass, from 0.1 dex at the low-mass end down to 0.05 dex at the high-mass end. As with our results mentioned earlier in this work, these findings are also qualitatively consistent with the study of \citet{Pasquali2010}. Taken at face value, our result suggests that the environment plays an important role in shaping the chemical enrichment histories of satellites of all masses, with satellites in more massive haloes tending to be more metal-rich than satellites in less massive haloes. 

\begin{figure}
\centering
\centerline{\includegraphics[width=\linewidth]{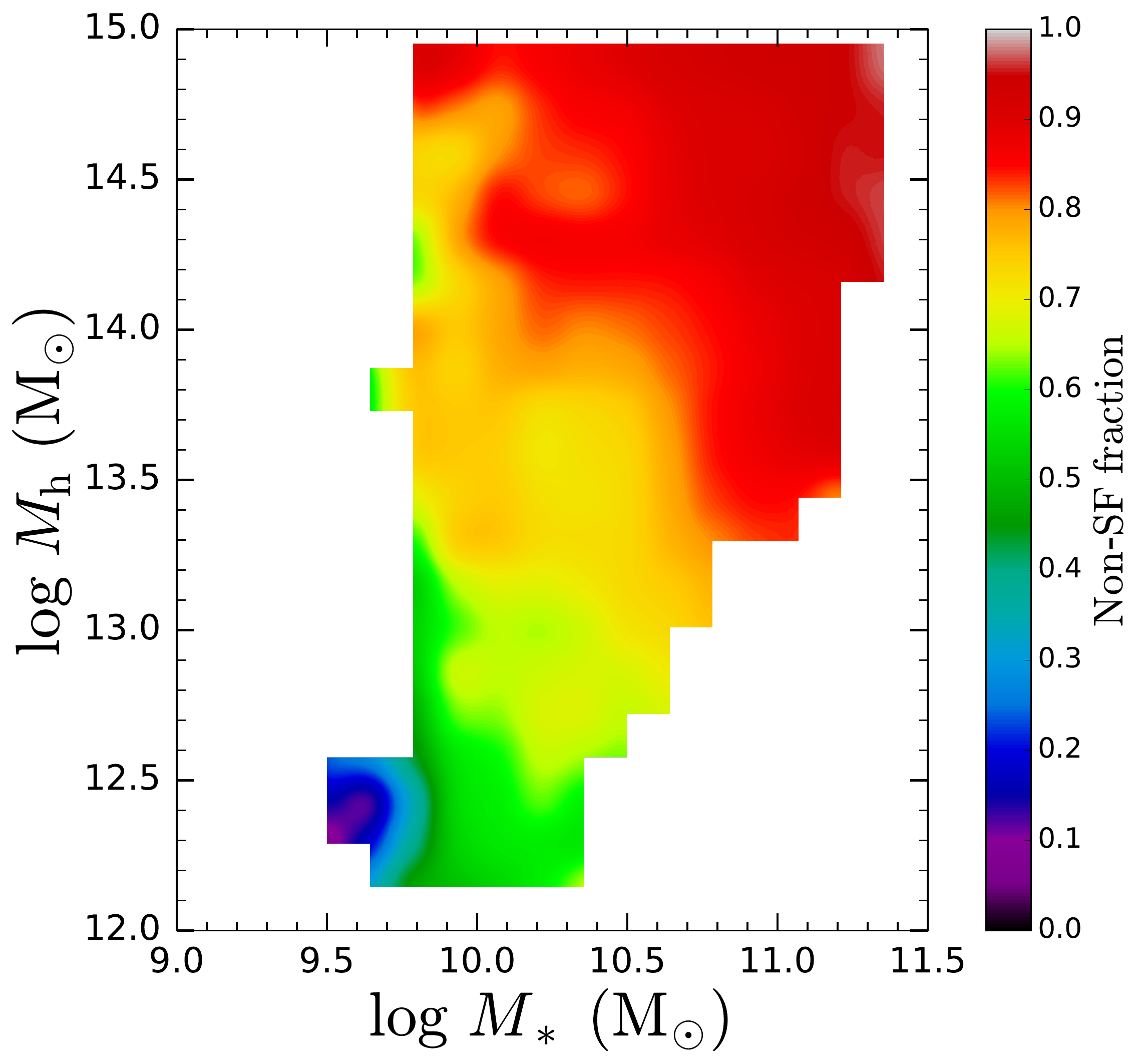}}
\caption{The non-SF fraction (i.e. the fraction of galaxies that are green valley or passive) for our sample of satellites (which has been biased by cuts on S/N), colour-coded as a function of stellar mass and halo mass.}
\label{fig:non_sf_fraction_hm_sat}
\end{figure}

However, as with the discussion of the stellar metallicity difference between the overall populations of centrals and satellites in Section~\ref{subsubsec:cent_sat_mzr}, it is important to bear in mind the quenched fraction effect when interpreting our results for the halo-mass dependence of the overall population of satellites in Fig.\@~\ref{fig:halo_mzr_mw_sat}. Numerous studies \citep[e.g.\@][]{Woo2013} have shown that, at a fixed stellar mass, the quenched fraction of satellites increases with increasing halo mass. For convenience we show this result in Fig.\@~\ref{fig:non_sf_fraction_hm_sat}, but displaying instead the dependence of the non-SF fraction on stellar mass and halo mass for our sample. Thus, the relative number of green valley and passive galaxies (which have higher stellar metallicities than star-forming galaxies of the same stellar mass) increases as one moves from the lower to the higher quartiles. As a result, the quenched fraction effect may be exaggerating the true, inherent dependence of the stellar metallicity of satellites on halo mass. 

We show the dependence of the mass-weighted stellar mass--stellar metallicity relation on halo mass for star-forming, green valley and passive satellites in the right, bottom-left and right panels, respectively. Note that the vertical scale and range is different for each panel. In general, we find that the trends with halo mass are less clear, smaller, and less significant than what was seen when studying the overall population of satellites. This may be the case for two reasons. First, the inherent dependence of stellar metallicity on halo mass is weak, so the trend for the overall population is perhaps mostly driven by the quenched fraction effect. Second, due to the lower statistics (since the data is split between star-forming, green valley and passive), the scaling relations have larger scatter, which complicates interpretation, and the error bars are larger, which makes it more difficult to distinguish between different quartiles. For star-forming satellites, we find that there is no clear trend with halo mass. The error bars on the stellar metallicities are relatively large because the dispersion in the stellar metallicities of star-forming galaxies is relatively high \citep[see][]{Trussler2020a}. For green valley satellites there is no trend at low stellar masses, but there is an indication that the stellar metallicity tends to increase with increasing halo mass at the high-mass end. Finally, for passive satellites, we find that there is a steady but small ($\sim$$0.04$~dex between the Low and High quartiles) increase of stellar metallicity with halo mass (of low significance) across the entire stellar mass range studied.

\subsubsection{Centrals} \label{subsubsec:sp_hm_cent}

\begin{figure*}
\centering
\centerline{\includegraphics[width=1\linewidth]{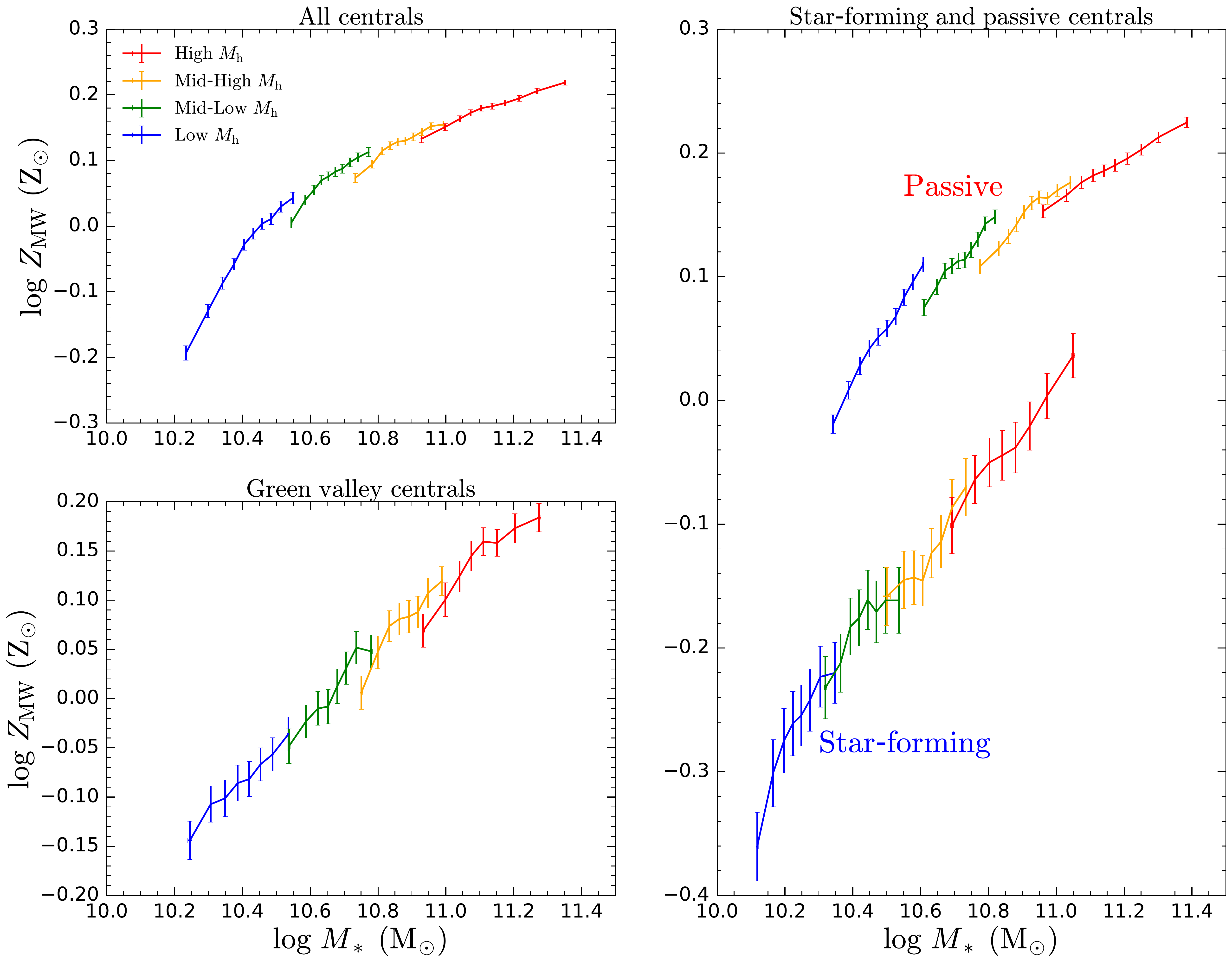}}
\caption{Similar to Fig.\@~\ref{fig:halo_mzr_mw_sat}, but now showing the mass-weighted stellar mass--stellar metallicity relation for centrals.}
\label{fig:halo_mzr_mw_cent}
\end{figure*}

We show the mass-weighted stellar mass--stellar metallicity relation for centrals in different halo mass quartiles in Fig.\@~\ref{fig:halo_mzr_mw_cent}. Given the correlation between the stellar mass of the central galaxy in a group and group halo mass \citep[see e.g.\@][]{Yang2008, Yang2009, Yang2012}, centrals in the higher halo mass quartiles tend to have higher stellar masses. Hence, there is little overlap in the stellar mass range covered between the different halo mass quartiles. In order to represent the trends with halo mass for centrals more clearly, we thus use a different stellar-mass binning procedure, now instead choosing mass bins such that there are an equal number of centrals per bin.

Interestingly, for the passive centrals in the right panel of Fig.\@~\ref{fig:halo_mzr_mw_cent}, there is an indication that, over the narrow range of stellar mass where there is overlap between subsequent quartiles, centrals of a given stellar mass are more metal-rich when they reside in lower mass haloes.
There is perhaps also a similar trend for green valley centrals in the bottom-left panel but it is much weaker than what is seen for passive centrals.

A possible explanation for this effect is that it may be driven by a potential progenitor bias, where the stellar metallicities of passive centrals of a given stellar mass may also depend on their formation histories, which in turn may depend on halo mass \citep[see][]{Man2019}.
Since the stellar metallicity of a passive central depends on both the stellar metallicity of its star-forming progenitor and the amount of chemical enrichment that takes place during quenching, any potential halo mass dependence in either of these two factors will determine the halo mass dependence of the stellar metallicity of the passive central. If, at a fixed stellar mass, passive centrals with larger halo masses formed earlier \citep[as suggested by][]{Man2019}, then their star-forming progenitors likely started quenching at a relatively higher redshift. Since the normalisation of the mass--metallicity relation decreases with redshift \citep[e.g.\@][]{Maiolino2008, Lian2018c}, then the stellar metallicities of the star-forming progenitors of passive centrals of a given stellar mass, but with larger halo masses, should be smaller. Conversely, however, since the typical gas fractions in galaxies increase with redshift \citep[e.g.\@][]{Schinnerer2016, Scoville2017, Tacconi2018}, the gas fractions of the star-forming progenitors in more massive haloes should be larger. As the amount of chemical enrichment that takes place during quenching depends on how many additional metal-rich stars are formed, which in turn depends on the size of the available gas reservoir \citep[and also the relative importance of different quenching mechanisms, namely starvation and outflows, for more details see][]{Peng2015, Trussler2020a}, the star-forming progenitors of passive centrals in more massive haloes likely undergo more chemical enrichment during quenching. Therefore, there are two competing and opposing effects (progenitor metallicity and enrichment during quenching), that both potentially depend on halo mass, that influence the stellar metallicity of the passive central. Depending on which effect is stronger, progenitor bias may or may not explain the observed trend with halo mass, i.e.\@ the vertical offset seen between subsequent halo mass quartiles in Fig.\@~\ref{fig:halo_mzr_mw_cent}.

\begin{figure*}
\centering
\centerline{\includegraphics[width=\linewidth]{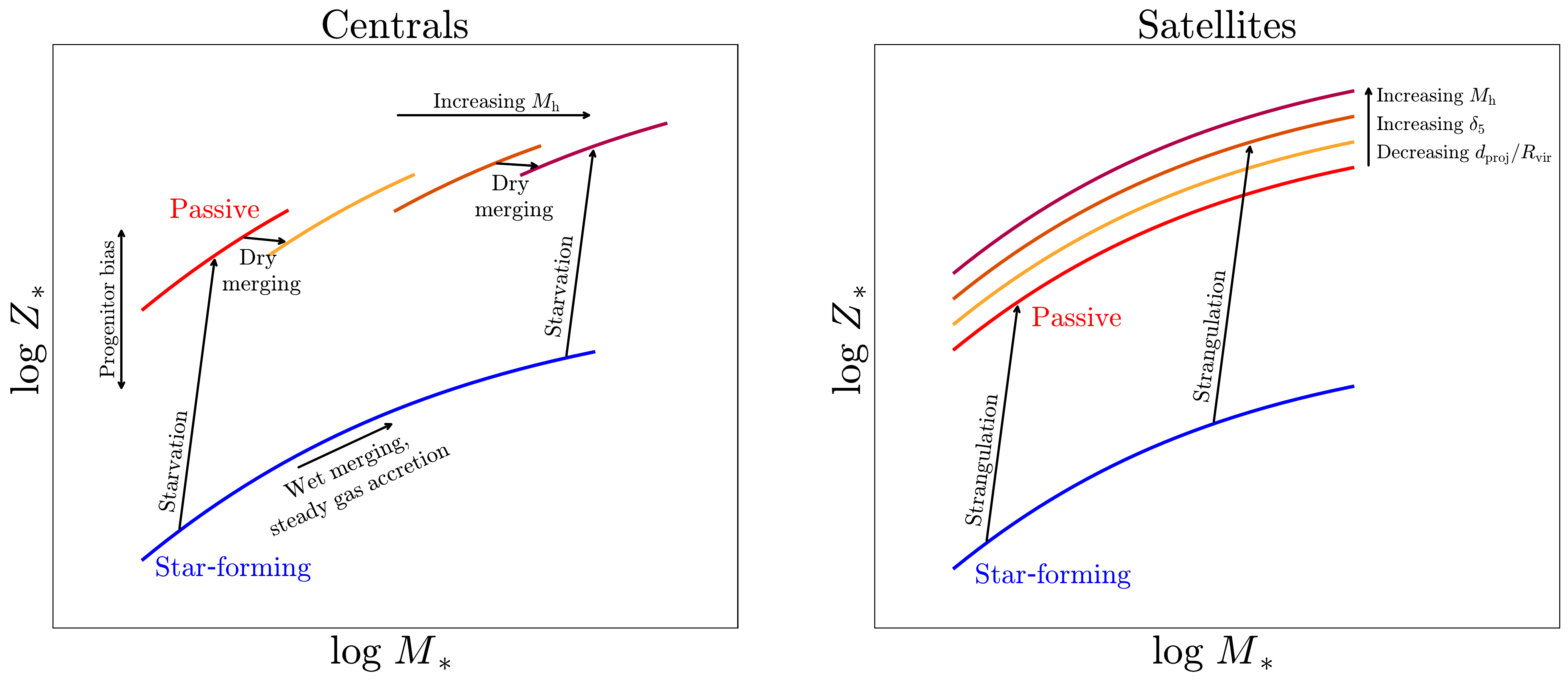}}
\caption{A schematic illustration showing how the evolution of galaxies in the stellar mass--stellar metallicity plane depends on environment. Left panel: The evolution of central galaxies is shown. Star-forming centrals (shown in blue) evolve along the Main Sequence, following a diagonal trajectory in the stellar mass--stellar metallicity plane, through steady accretion of gas from the cosmic web and/or by acquiring gas in short bursts from wet mergers. Passive centrals, split into quartiles in halo mass (shown in various shades of red), evolve through dry mergers, following a $\sim$horizontal/slightly downward trajectory in the stellar mass--stellar metallicity plane, where both their stellar mass ($M_*$) and halo mass ($M_\mathrm{h}$) increase (causing them to move to the next halo mass quartile), but their stellar metallicity ($Z_*$) remains relatively unchanged/decreases slightly as the satellite galaxy involved in the merger is, by definition, of a lower stellar mass and therefore has, on average, a lower stellar metallicity. Alternatively, a potential progenitor bias may also explain the trend with halo mass, where, at a fixed stellar mass, passive centrals in more massive haloes may potentially have lower stellar metallicities because they potentially quenched earlier. Given the correlation between the stellar mass of central galaxies and their group halo mass, there is limited overlap in the stellar mass range covered between the different halo mass quartiles. Star-forming centrals tend to quench through starvation to become passive centrals. Right panel: The evolution of satellite galaxies is shown. The stellar metallicity of star-forming satellites does not show much dependence on halo mass ($M_\mathrm{h}$), overdensity ($\delta_5$) or projected distance from their central ($d_\mathrm{proj}/R_\mathrm{vir}$). In contrast, the stellar metallicity of passive satellites tends to increase with increasing halo mass, increasing overdensity and with decreasing projected distance. Given that environmental effects may be contributing to the starvation of satellites, we label the transition from star-forming to passive as being driven by strangulation.}
\label{fig:schematic_cents_sats_evolution}
\end{figure*}

Alternatively, rather than being offset vertically (i.e.\@ in metallicity), the halo mass quartiles for passive
central galaxies may in fact be offset horizontally (i.e.\@ in stellar mass). This scenario is depicted more
clearly in the left panel of Fig.\@~\ref{fig:schematic_cents_sats_evolution}, which shows a schematic illustration
for the evolution of central galaxies. Passive centrals, due to their lack of gas, subsequently evolve through dry
mergers, where both their stellar mass and halo mass increase (causing them to move to the next halo mass
quartile), but their stellar metallicity remains relatively unchanged. In fact, the stellar metallicity of the passive central will slightly decrease as it merges with another galaxy, as that galaxy must have a lower stellar mass than the central (otherwise the central would be a satellite, by definition) and thus, on average, has a lower stellar metallicity. Thus, passive central galaxies evolve along a $\sim$horizontal/slightly downward trajectory on the stellar mass--stellar metallicity plane, with the normalisation of the stellar
mass--stellar metallicity relation becoming progressively lower for the higher halo mass quartiles. In contrast,
star-forming centrals evolve along the star-forming Main Sequence, following a diagonal trajectory in the stellar
mass--stellar metallicity plane as they steadily accrete gas from the cosmic web and/or acquire gas in short bursts
from wet mergers, causing both their stellar mass and stellar metallicity to increase as progressively more and
more metal-rich stars are formed out of the gas reservoir. In addition, as shown in \citet{Peng2015} and
\citet{Trussler2020a}, and depicted in Fig.\@~\ref{fig:schematic_cents_sats_evolution}, the large difference in
chemical enrichment between star-forming and passive galaxies necessitates that most star-forming galaxies quench
through an extended phase of starvation (a shut down of cold gas accretion). In addition, as discussed in \citet{Trussler2020a}, while starvation is likely to be the prerequisite of quenching, outflows can also
contribute, with the combination of starvation and outflows being responsible for quenching the majority of galaxies.

In the right panel of Fig.\@~\ref{fig:schematic_cents_sats_evolution} we show a schematic illustration for the
evolution of satellite galaxies. The stellar metallicity of star-forming satellites have little dependence
on halo mass (as already shown), as well as
on overdensity or projected distance (as we will see later in this Section).
Again, star-forming satellites typically quench through starvation to become passive satellites, though, given that
environmental effects may be at play, we label this as `strangulation' in this case.
The stellar metallicity of passive satellites tends to increase with halo mass, as well as with increasing overdensity and decreasing projected distance from their central (as will be shown later in this Section).

\subsection{Overdensity} \label{subsec:sp_od}

Given the weak environmental trends found using halo masses in the previous section, we now investigate trends with alternative environmental parameters to verify whether a consistent but weak environmental dependence emerges from our analysis. In this section we study the dependence of the mass-weighted stellar mass--stellar metallicity relation on the local overdensity, by binning galaxies in quartiles of the local overdensity. 

\subsubsection{Satellites}

\begin{figure*}
\centering
\centerline{\includegraphics[width=1\linewidth]{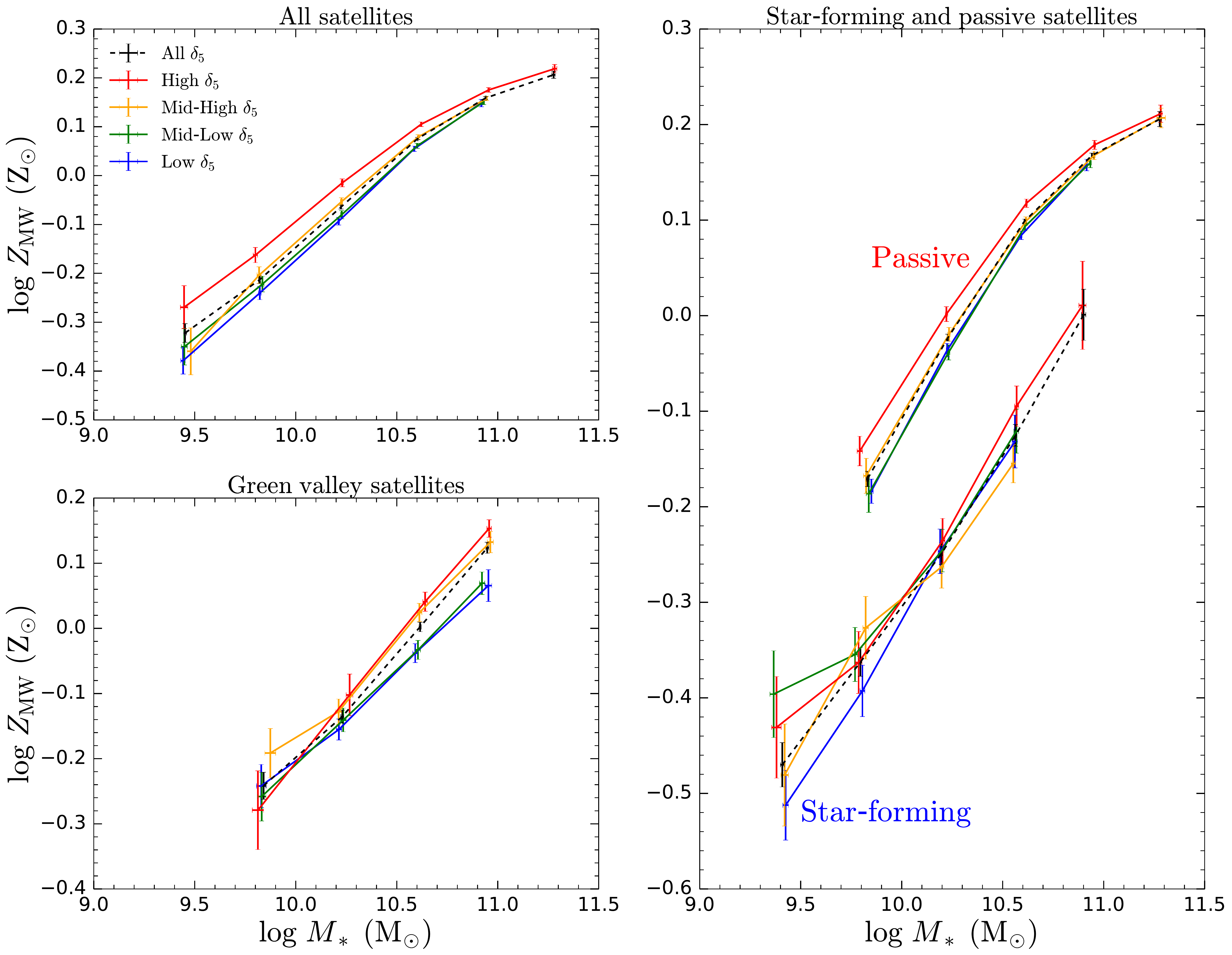}}
\caption{Similar to Fig.\@~\ref{fig:halo_mzr_mw_sat}, but now binning in quartiles of overdensity $\delta _5$.}
\label{fig:od_mzr_mw_sat}
\end{figure*}

We show the mass-weighted stellar mass--stellar metallicity relation for satellites in different overdensity quartiles in Fig.\@~\ref{fig:od_mzr_mw_sat}. While the stellar metallicity of the overall population of satellites (top-left panel) tends to increase with increasing overdensity, we find a much weaker environmental dependence once star-forming (right panel, no trend), green valley (bottom-left, weak trend) and passive satellites (right panel, weak trend) have been separated. Qualitatively, our findings for the overdensity analysis are very similar to what was found in the halo mass analysis. Given the correlation between halo mass and overdensity \citep[see][]{Woo2013}, the similarity between our results using halo masses and overdensities is to be expected.

Interestingly, most of the variation of stellar metallicity with overdensity for the overall population, green valley and passive satellites comes about from the jump to the Mid-High and High quartiles, with little to no variation between the Low and Mid-Low quartiles. We suspect that this may be due to the manner in which the overdensity is calculated. Small overdensities typically correspond to galaxies that are in groups with membership $\leq 5$, (i.e.\@ the $N = 5$ used in the calculation of the overdensity $\delta _5$), for which the fifth nearest neighbour is in another group. Hence the overdensity is not a strong tracer of environmentally-driven processes in this regime, as it characterises the density of galaxies between groups, rather than the density of galaxies within groups. \citep[for more details, see][]{Woo2013}.

\subsubsection{Centrals}

\begin{figure*}
\centering
\centerline{\includegraphics[width=1\linewidth]{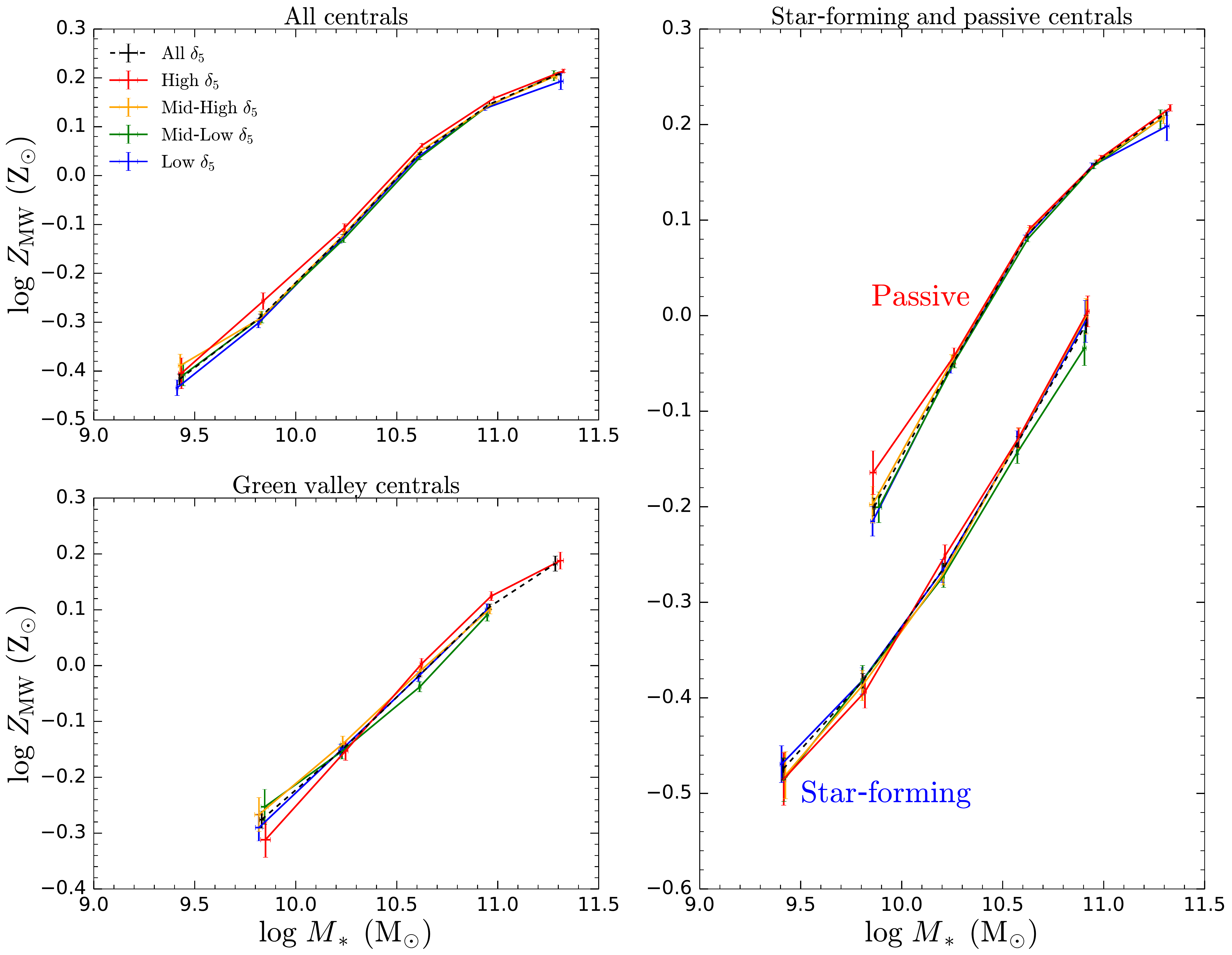}}
\caption{Similar to Fig.\@~\ref{fig:od_mzr_mw_sat}, but now showing the mass-weighted stellar mass--stellar metallicity relation for centrals.}
\label{fig:od_mzr_mw_cent}
\end{figure*}

We show the mass-weighted stellar mass--stellar metallicity relation for centrals in different quartiles of overdensity in Fig.\@~\ref{fig:od_mzr_mw_cent}. We find, for the overall population of centrals in the top-left panel, that the stellar metallicity tends to increase slightly with increasing overdensity, but this is primarily due to the jump to the High $\delta_5$ quartile. Since the quenched fraction of centrals increases weakly with overdensity \citep[see e.g.\@][]{Woo2013}, this is likely to be driven by the quenched fraction effect. Indeed, we do not find any clear trend between stellar metallicity and overdensity for star-forming, green valley or passive centrals. This lack of environmental dependence is consistent with the works of \citet{Thomas2010} and \citet{Goddard2017a}, who also find no correlation between stellar metallicity (or stellar metallicity gradients) and overdensity.

\subsection{Projected distance and summary} \label{subsec:sp_proj}

Finally, we study the dependence of the mass-weighted stellar mass--stellar metallicity relation on projected distance from the central galaxy (normalised by the virial radius of the galaxy group), by binning satellites in quartiles of projected distance. Our study is motivated by the fact that some physical processes, like e.g.\@ ram pressure stripping, may only become active close to the centres of groups/clusters (i.e.\@ at small projected distances), or that sufficient time may need to elapse following the infall of a satellite into a group/cluster before the impact of these processes becomes apparent (again, at small projected distances).

\begin{figure*}
\centering
\centerline{\includegraphics[width=1\linewidth]{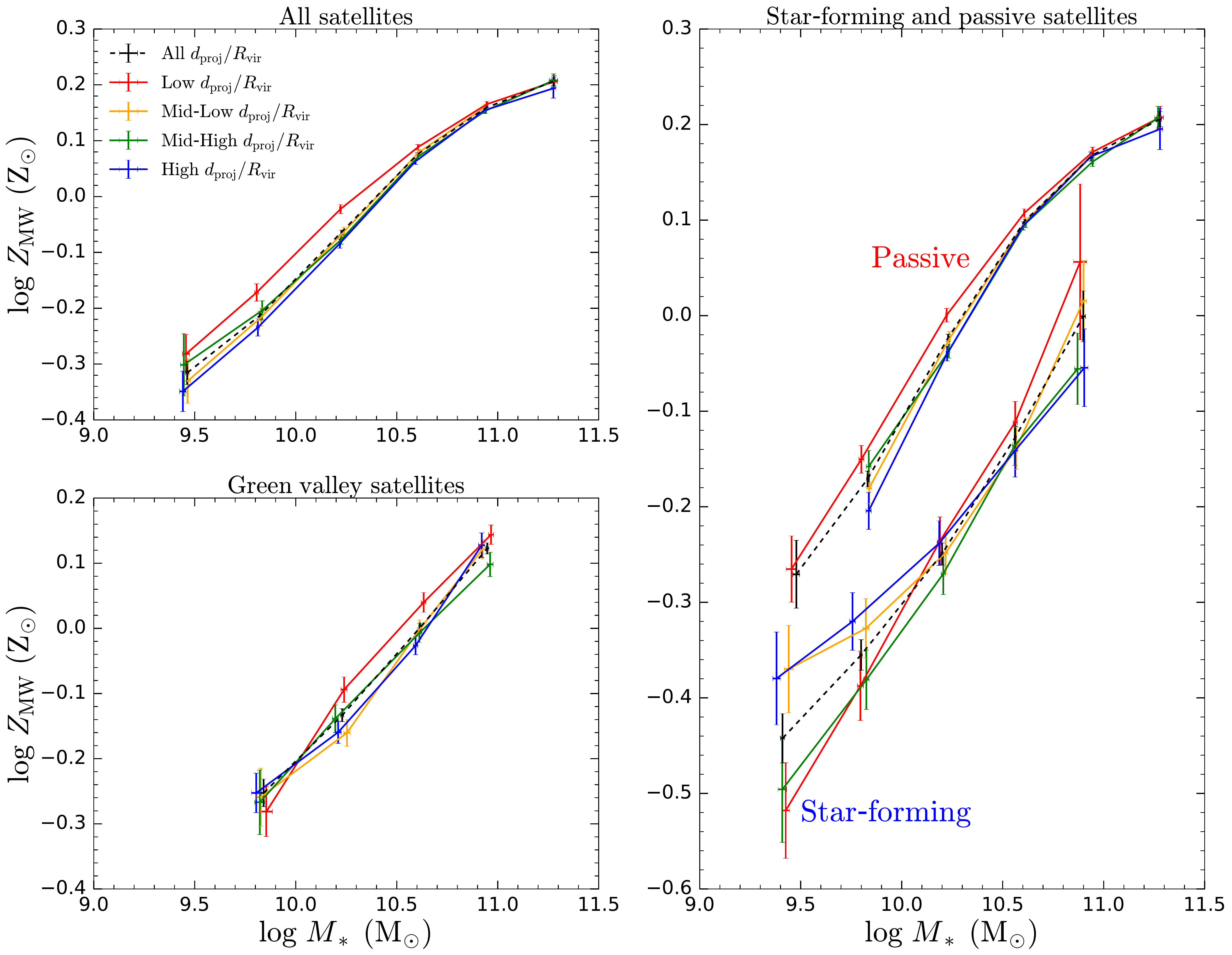}}
\caption{Similar to Fig.\@~\ref{fig:halo_mzr_mw_sat}, but now binning in quartiles of projected distance $d_\mathrm{proj}/R_\mathrm{vir}$. Note that the order of the colour scheme for the quartiles has been reversed. The Low, Mid-Low, Mid-High and High quartiles are now coloured red, orange, green and blue, respectively.}
\label{fig:proj_mzr_mw_sat}
\end{figure*}

We show the mass-weighted stellar mass--stellar metallicity relation for satellites in different projected distance quartiles in Fig.\@~\ref{fig:proj_mzr_mw_sat}. While the stellar metallicity of the overall population of satellites (shown in the top-left panel) tends to increase with decreasing projected distance to the central galaxy of the group, we find a much weaker environmental dependence once star-forming (right panel, no trend), green valley (bottom-left panel, weak trend) and passive satellites (right panel, weak trend) have been separated. These results are qualitatively consistent with what was seen in the halo mass and overdensity analysis. This might be because, for groups with more than five galaxies (i.e.\@ more than the $N=5$ we use in the calculation of the overdensity $\delta _5$), overdensity anti-correlates with projected distance \citep[for more details, see][]{Woo2013}.

Thus, to summarise, in Sections~\ref{subsec:sp_hm}, \ref{subsec:sp_od} and \ref{subsec:sp_proj}, we investigated the dependence of the stellar mass--stellar metallicity relation on further environmental parameters: halo mass, local overdensity and (for satellites only) projected distance from the central galaxy, respectively. We obtained qualitatively consistent results across these three environmental parameters. We found that the stellar metallicities of star-forming satellites do not depend on environment (see Figs.\@~\ref{fig:halo_mzr_mw_sat}, \ref{fig:od_mzr_mw_sat} and \ref{fig:proj_mzr_mw_sat}). In contrast, the stellar metallicities of passive and green valley satellites increase weakly with the density of the environment within which they reside, i.e.\@ with increasing halo mass (see Fig.\@~\ref{fig:halo_mzr_mw_sat}), increasing local overdensity (see Fig.\@~\ref{fig:od_mzr_mw_sat}) and decreasing projected distance from their central galaxy (see Fig.\@~\ref{fig:proj_mzr_mw_sat}). Furthermore, we found that the stellar metallicities of central galaxies do not depend on the local overdensity (see Fig.\@~\ref{fig:od_mzr_mw_cent}). Finally, we found a unique feature in the stellar mass--stellar metallicity relation for passive centrals, where galaxies in more massive haloes have larger stellar mass at constant stellar metallicity (see Fig.\@~\ref{fig:halo_mzr_mw_cent}). These key environmental trends are summarised in Fig.\@~\ref{fig:schematic_cents_sats_evolution}. \\
\indent We compare the environmental trends that we have found for stellar metallicities against trends that previous studies have found for \textit{gas-phase} metallicities in Appendix~\ref{sec:discussion}.

\section{Environmental quenching} \label{sec:env_quenching}

Our analysis thus far has focussed on the stellar mass--stellar metallicity relation. We now shift our attention to
the stellar metallicity difference between star-forming and passive galaxies to investigate how the
environment contributes to the quenching of galaxies. As discussed in detail in \citet{Peng2015} and
\citet{Trussler2020a}, the stellar metallicity difference between star-forming and passive galaxies can be used to
distinguish between different quenching mechanisms as the amount of chemical enrichment during the quenching phase
depends on the mechanism. In the case of a galaxy quenching by sudden gas removal (e.g.\@ outflows or rapid
ram pressure),
the stellar metallicity increase $\Delta Z_*$ during the quenching phase is small as relatively few additional stars are formed out of the available gas reservoir. On the other hand, galaxies quenching through starvation (the halting of cold gas accretion) no longer have their ISM diluted by the inflow of pristine gas and form relatively many additional metal-rich stars, so the increase in stellar metallicity during the quenching phase is large. Thus, if galaxies quench primarily through starvation, the stellar metallicity difference between star-forming and passive galaxies of the same stellar mass should be large. In contrast, if galaxies quench primarily through sudden gas removal, the stellar metallicity difference should be small.\footnote{In reality, galaxies are likely to quench through a combination of starvation and gas removal processes, as indicated in \citet{Trussler2020a}.}

We will use the stellar metallicity difference between star-forming and passive galaxies as a proxy for the
prevalence of starvation as a mechanism for quenching star formation in galaxies. In particular, we wish to
investigate whether environmental effects contribute to the starvation of galaxies, and if so, to put constraints
on the physical origin of any environmentally-driven starvation, using the chemical evolution model of \citet{Trussler2020a}. We will study environmental quenching through a comparison between
the stellar metallicity difference for centrals and satellites in different quartiles of halo mass,
overdensity and projected distance in Section~\ref{subsec:quench_env}. We discuss both the qualitative and quantitative constraints on environmental quenching in Section~\ref{subsec:quench_constraints}.

\subsection{Stellar metallicity differences} \label{subsec:quench_env}

In this section we study how the stellar metallicity difference between star-forming and passive galaxies depends on halo mass, overdensity and projected distance, for both centrals and satellites.

In Section~\ref{sec:env_dependence} we found a subtle indication that the stellar metallicity of both green valley
and passive satellites depends on environment, with the stellar metallicities tending to increase slightly with both
increasing halo mass and overdensity, and with decreasing projected distance. In contrast, we found no trend between stellar metallicity and environment for star-forming satellites, neither for halo mass,
nor for overdensity or projected distance. Thus, in this section,
we will make the assumption that the stellar metallicity of star-forming satellites (and star-forming centrals) does
not depend on environment, i.e.\@ we will not bin the star-forming satellite population into quartiles and instead
use the relation for the entire population of star-forming satellites (given by the curves named `All' in Figs.\@~\ref{fig:halo_mzr_mw_sat}, \ref{fig:od_mzr_mw_sat} and \ref{fig:proj_mzr_mw_sat}). 

\begin{figure*}
\centering
\centerline{\includegraphics[width=1\linewidth]{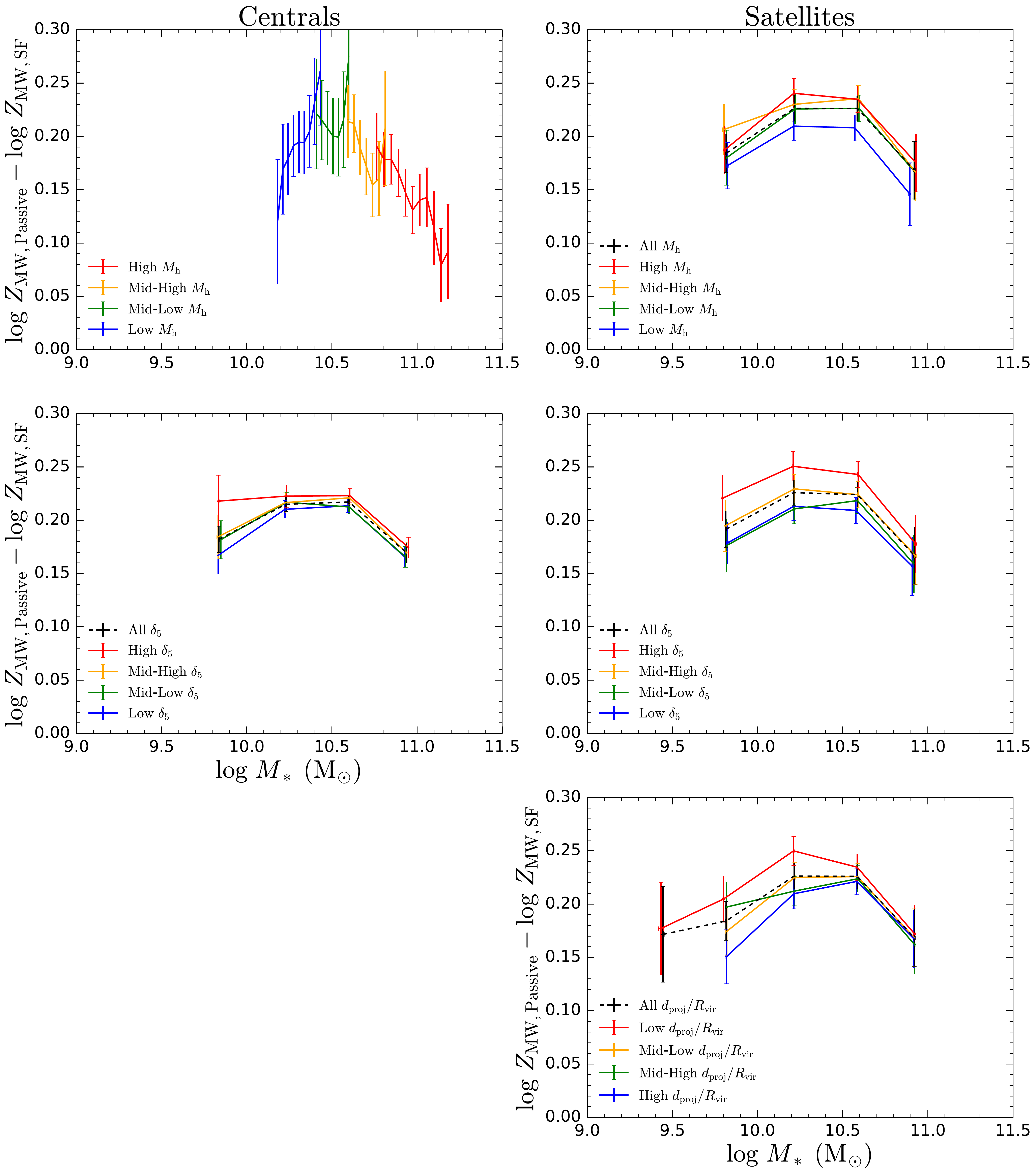}}
\caption{The mass-weighted stellar metallicity difference $\Delta \log Z_\mathrm{MW} = \log Z_\mathrm{MW, Passive} - \log Z_\mathrm{MW, SF}$ between star-forming and passive galaxies for centrals (left panels) and satellites (right panels) in quartiles of different environmental measures. The Low (blue), Mid-Low (green), Mid-High (orange) and High (red) quartiles correspond to the $0$$^\mathrm{th}$--$25$$^\mathrm{th}$, $25$$^\mathrm{th}$--$50$$^\mathrm{th}$, $50$$^\mathrm{th}$--75$^\mathrm{th}$ and $75$$^\mathrm{th}$--$100$$^\mathrm{th}$ percentiles, respectively. We reverse the order of the colour scheme for the projected distance quartiles. As we found that the stellar metallicities of star-forming satellites (and star-forming centrals) do not depend on environment, we only use quartiles for the passive satellite (and passive central) population. We also show the stellar metallicity differences obtained without splitting into quartiles (named All, shown in dashed black). Top panels: passive galaxies are split into quartiles of group halo mass $M_\mathrm{h}$. Middle panels: passive galaxies are split into quartiles of overdensity $\delta_5$. Bottom panel: passive satellites are split into quartiles of projected distance from their central $d_\mathrm{proj}/R_\mathrm{vir}$.}
\label{fig:mz_hm_od_proj_qg}
\end{figure*}

We show how the mass-weighted stellar metallicity difference between star-forming and passive galaxies depends on
group halo mass (top panels), overdensity (middle panels) and projected distance (bottom panel) for both centrals
(left panels) and satellites (right panels) in Fig.\@~\ref{fig:mz_hm_od_proj_qg}. 

For centrals, we find that the stellar metallicity difference shows no dependence on environment. There is perhaps a subtle (but not significant) indication that the difference increases with overdensity, but this is only a very marginal effect at most. 

For satellites, there is a relatively stronger (but still not significant) trend between the stellar metallicity
difference and environment, with the stellar metallicity difference tending to increase (by $<0.05$~dex) with the density of the environment within which satellites reside (i.e.\@ increasing halo mass and overdensity, and decreasing projected distance). Despite
the excellent overall statistics in SDSS DR7, there are only a relatively small number of satellites (especially with
respect to centrals) in our sample. Hence, although the trends for satellites appear to be stronger than for
centrals, these trends are still only marginally significant: at most at the $2\sigma$ level in some mass bins. Indeed, given the relatively weak inherent dependence of stellar metallicity on environment (see Section~\ref{sec:env_dependence}), a study of stellar metallicity differences (which have larger errors) is likely to only find marginally significant trends even if the statistics are high. 

\subsection{Constraints on environmental quenching} \label{subsec:quench_constraints}

As discussed in \citet{Peng2015} and \citet{Trussler2020a}, and also outlined at the beginning of this section, the large stellar metallicity difference observed
between the overall population of star-forming and passive galaxies provides strong evidence indicating that, for the majority of galaxies, quenching likely involved an extended
phase of starvation, with gas removal through outflows or ram pressure stripping also contributing to the quenching. This starvation process is generally ascribed to two primary classes of phenomena: i) mass-dependent phenomena, in
which the halo of the galaxy is heated by AGN feedback (via energy injection through jets or winds) or through
gravitational shock-heating, resulting in the suppression of cold accretion onto the galaxy
\citep[e.g.\@][]{Dekel2006, Croton2006,Fabian2012,Brownson2019};
ii) environmental effects, often referred to as `strangulation', that occur when satellite galaxies plunge into the hot
halo of a larger galaxy or an overdense region (i.e.\@ a group or cluster) and have their circumgalactic medium stripped and/or become disconnected from the cosmic web, which prevents the satellite galaxy from accreting further fresh gas
\citep[e.g.\@][]{Larson1980, VanDenBosch2008, Feldmann2015a, VandeVoort2017b, AragonCalvo2019}. \\
\indent The lack of environmental dependence in the stellar metallicity difference between star-forming and passive central galaxies
indicates, not surprisingly, that the starvation responsible for quenching central galaxies is not driven
by environmental phenomena, but primarily by mass-related phenomena such as halo heating by AGN and halo gravitational shock heating, or the recently proposed angular momentum quenching \citep{Peng2020, Renzini2020}, where a galaxy can be deprived of new molecular gas when the inflowing gas accreted from the IGM comes in with excessive angular momentum. \\
\indent The presence of environmental effects (although only at the 2$\sigma$ level) on the stellar metallicity difference
between star-forming and passive satellite galaxies indicates that environmental `strangulation' does play
a role in the starvation process that quenches satellite galaxies. However, since this environmental effect is small ($<0.05$~dex) relative to the overall stellar metallicity difference between star-forming and passive satellite galaxies \citep[$0.10$--$0.22$~dex, see also][]{Trussler2020a}, the bulk of the starvation of satellite galaxies is still due to mass-related effects. \\
\indent We investigate environmental quenching more quantitatively by using the chemical evolution model of \citet{Trussler2020a}. This model derives valuable constraints on parameters associated with quenching, by aiming to simultaneously reproduce both the high stellar metallicities (see e.g.\@ Fig.\@~\ref{fig:mzr_mw_cent_sat}) and the low $\mathrm{SFR}$s of local passive galaxies \citep[see][]{Trussler2020a}, given the properties of their high-redshift star-forming progenitors. In particular, the \citet{Trussler2020a} model utilises the mass- and environment-dependent stellar metallicities, stellar ages and SFRs of local star-forming and passive galaxies to derive the epoch associated with the onset of quenching $z_\mathrm{q}$, quenching time-scales (both the duration $t_\mathrm{quench}$ and $e$-folding time $\tau_\mathrm{q}$) and the relative importance of starvation and gas removal processes (i.e.\@ outflows and/or gas stripping) in driving galaxy quenching. We assume that galaxies quench through a combination of starvation and gas removal processes, i.e.\@ we set the gas accretion rate $\Phi = 0$, and gas is ejected from the galaxy at a net ejection rate $\Lambda = \lambda_\mathrm{eff} \times \mathrm{SFR}$, where $\lambda_\mathrm{eff}$ is the `effective'  (i.e.\@ the expelled gas is not reaccreted) mass-loading factor. \\
\begin{figure*}
\centering
\centerline{\includegraphics[width=1\linewidth]{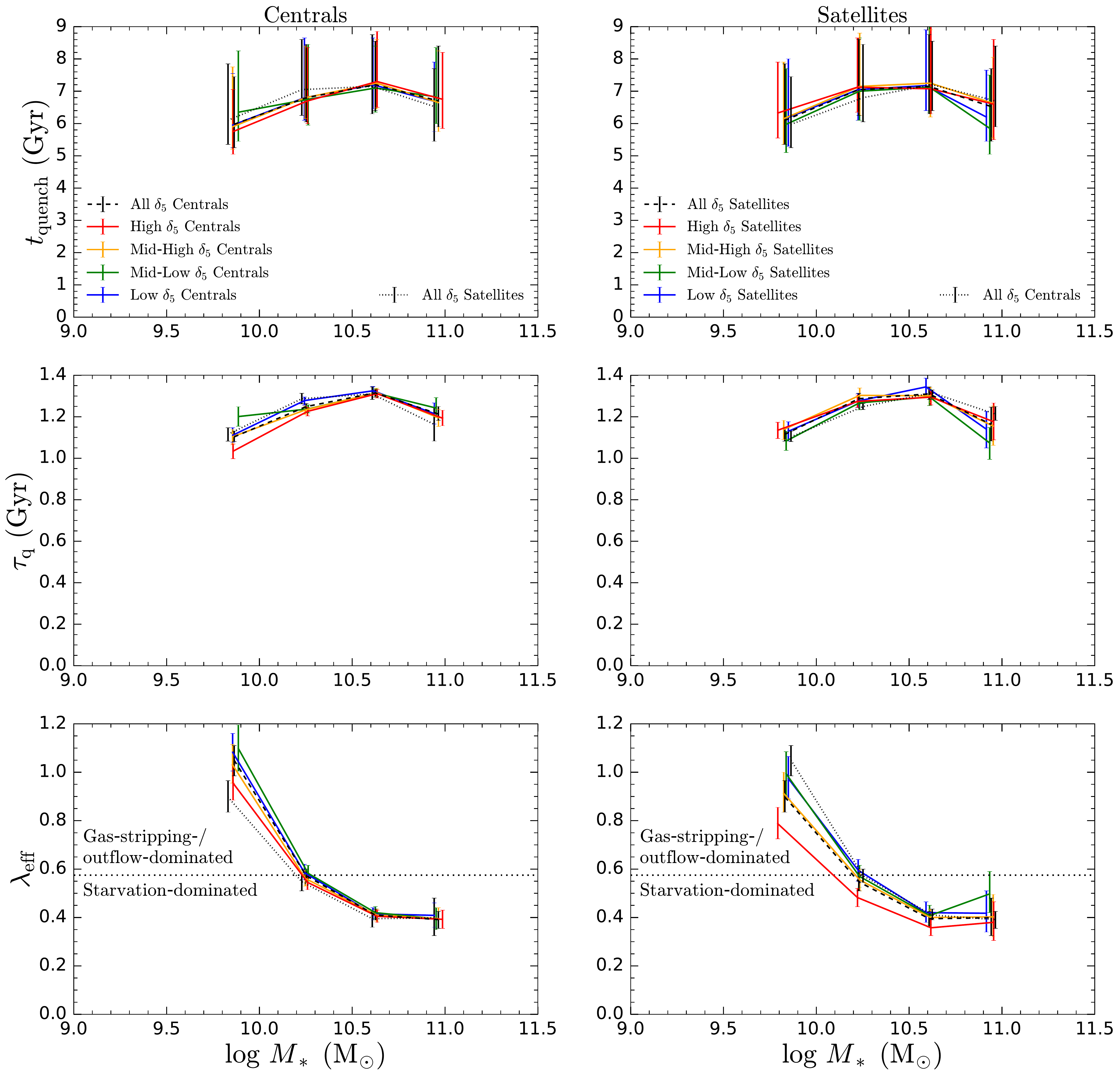}}
\caption{Quantitative constraints on environmental quenching derived by applying the \citet{Trussler2020a} chemical evolution model to the mass- and environment-dependent (i.e.\@ for both centrals and satellites, and as a function of local overdensity $\delta_5$) mass-weighted stellar metallicity differences $\Delta \log Z_\mathrm{MW} = \log Z_\mathrm{MW, Passive} - \log Z_\mathrm{MW, SF}$ between star-forming and passive galaxies in the middle panel of Fig.\@~\ref{fig:mz_hm_od_proj_qg}. Results for centrals and satellites are shown in the left and right panels, respectively. The colour coding and labelling is identical to the middle panel of Fig.\@~\ref{fig:mz_hm_od_proj_qg}. In order to facilitate an easier comparison between centrals and satellites, we also show (in dotted black) the results obtained when centrals and satellites are not split into quartiles of environment (i.e.\@ corresponding to the curves named `All' in Fig.\@~\ref{fig:mz_hm_od_proj_qg}) in the right and left panels, respectively. Top panels: The durations of quenching $t_\mathrm{quench}$, which specify how long it takes for a galaxy to transition from star-forming to passive, as a function of stellar mass $M_*$. Middle panels: The $e$-folding time-scales for quenching $\tau_\mathrm{q}$, which indicate the time-scale over which the $\mathrm{SFR}$ decreases by a factor of $e$. Bottom panels: The `effective' mass-loading factors $\lambda_\mathrm{eff}$ (which describe the net gas expulsion rate during quenching) required to simultaneously reproduce the mass- and environment-dependent stellar metallicities $Z_*$, and mass-dependent $\mathrm{SFR}$s of local passive galaxies, given the properties of their high-redshift star-forming progenitors. The horizontal dotted line is the $\lambda_\mathrm{eff}$ value for which the net gas expulsion rate ($\lambda_\mathrm{eff}\Psi$) is the same as the net gas consumption rate from star formation. Galaxies with $\lambda_\mathrm{eff}$ above (below) this boundary are considered to be gas-stripping-/outflow-dominated (starvation-dominated).}
\label{fig:model_qg_env_od}
\end{figure*}
\indent The quantitative constraints on environmental quenching derived using the \citet{Trussler2020a} chemical evolution model are shown for both centrals (left panels) and satellites (right panels) in Fig.\@~\ref{fig:model_qg_env_od}. In order to facilitate an easier comparison between centrals and satellites, we also show (in dotted black) the results obtained when centrals and satellites are not split into quartiles of environment (i.e.\@ corresponding to the curves named `All' in Fig.\@~\ref{fig:mz_hm_od_proj_qg}) in the right and left panels, respectively. Our analysis henceforth will focus on environmental trends with the local overdensity $\delta_5$, though we note that we obtain very similar results for satellites when halo masses $M_\mathrm{h}$ and projected distances $d_\mathrm{proj}/R_\mathrm{vir}$ are used as measures of environment instead. 

The durations of quenching $t_\mathrm{quench}$ derived using our chemical evolution model are shown in the top panels of Fig.\@~\ref{fig:model_qg_env_od}. We find that $t_\mathrm{quench}$ is roughly mass-independent at 6--7~Gyr. Interestingly, we also find that $t_\mathrm{quench}$ shows no environmental dependence, as neither centrals nor satellites show any trends with overdensity, and furthermore, the quenching time-scales for centrals and satellites are comparable at all masses. 

These environmental results are qualitatively consistent with some earlier studies of environmental quenching who also found no environmental dependence for the quenching time-scales \citep[e.g.\@][]{Wetzel2013a, Balogh2016, Fossati2017}. 
 
We show the $e$-folding time-scales for quenching $\tau_\mathrm{q}$ in the middle panels of Fig.\@~\ref{fig:model_qg_env_od}. We find that $\tau_\mathrm{q}$ is roughly mass-independent at $\sim$1.2~Gyr. Furthermore, we find that $\tau_\mathrm{q}$ shows no environmental dependence. Neither centrals nor satellites show any trends with overdensity, and the $\tau_\mathrm{q}$ values for centrals and satellites are comparable at all masses. 
 
Finally, we show the `effective' mass-loading factors $\lambda_\mathrm{eff}$ during quenching in the bottom panels of Fig.\@~\ref{fig:model_qg_env_od}. We find that $\lambda_\mathrm{eff}$ decreases with increasing stellar mass, with `effective' outflows and gas stripping (which are capable of permanently removing gas from the galaxy) playing an increasingly important role in quenching low-mass galaxies. Henceforth, we will refer to galaxies for which the net gas expulsion rate ($\lambda_\mathrm{eff}\Psi$) is larger (smaller) than the net gas consumption rate from star formation as gas-stripping-/outflow-dominated (starvation-dominated).

As shown in the bottom-left panel, we find that the mass-loading factors $\lambda_\mathrm{eff}$ for central galaxies show no environmental dependence, as the $\lambda_\mathrm{eff}$ values for centrals in high density (shown in red) and low density (blue) environments are comparable at all masses. In contrast, the mass-loading factors for satellite galaxies (at all stellar masses) clearly decrease with the density of the environment within which they reside (i.e. increasing $\delta_5$), with the quenching of satellite galaxies in high density environments being relatively more starvation-dominated than for satellite galaxies of the same stellar mass in low-density environments (which are relatively more gas-stripping-/outflow-dominated). Furthermore, the strength of this environmental effect decreases in size (both linearly and logarithmically) and in significance with increasing stellar mass, with the offset between the High (shown in red) and Low (blue) quartiles decreasing from $3\sigma$ at $M_* = 10^{9.8}~\mathrm{M}_\odot$ to $< 0.5\sigma$ at $M_* = 10^{11}~\mathrm{M}_\odot$. \\
\indent The aforementioned environmental trends for the mass-loading factor $\lambda_\mathrm{eff}$ are further highlighted by comparing centrals against satellites. We find that satellites have smaller $\lambda_\mathrm{eff}$ than centrals at all stellar masses, with the offset between centrals and satellites decreasing with increasing stellar mass. \\
\indent Thus, our results indicate that it is in fact starvation, rather than gas stripping, that becomes progressively more important in quenching star formation in satellites in progressively denser environments. Furthermore, while environmental effects contribute moderately to the starvation and/or confinement of gas in low-mass satellite galaxies in dense environments, environmental effects only play a minor role in the evolution of massive satellites and satellites in low-density environments. \\
\indent We note that previous observational studies have provided direct evidence of environmentally-driven gas stripping processes in action in the local Universe, with both imaging and spectroscopic analyses confirming that satellite galaxies can have their gas stripped (into tails) as they plunge into the dense environments of galaxy groups and clusters \citep[see e.g.\@][]{Merluzzi2013, Fumagalli2014, Fossati2016a}. These observational results may perhaps be taken to indicate that gas stripping processes (rather than starvation) become increasingly important in quenching $z=0$ satellites as the density of the environment within which satellites reside increases. However, our analysis suggests that these observations may in fact present a somewhat biased view of the role of gas stripping processes in satellite quenching, as direct observational evidence for gas accretion (or a lack thereof, i.e.\@ starvation) onto galaxies is still rare \citep[see e.g.\@][]{Sancisi2008, Bouche2013}, due to the technical challenge in detection. Indeed, other works have suggested that satellites predominantly quench due to a shutdown of cosmological gas accretion, with gas stripping processes only playing a relatively minor role, especially beyond $z=0$ \citep[see e.g.\@][]{Balogh2016, Brown2017b, Fossati2017}. While environmental processes may be contributing to the stripping of gas in satellite galaxies, our results suggest that the environment contributes even more strongly to the halting of cold gas accretion onto satellites and/or the confinement (i.e.\@ enhanced recycling) of outflowing gas in satellites \citep[see e.g.\@][]{Bahe2012, Pasquali2012}, with the quenching of satellite galaxies becoming progressively more starvation-dominated (relative to gas-stripping-/outflow-dominated) as the density of the environment within which they reside increases.

\section{Summary and conclusions} \label{sec:conclusions}
We have utilised the statistical power of SDSS DR7 to investigate the environmental dependence of the stellar populations of galaxies, as well as environmental quenching. We obtain the following results:
\begin{enumerate}
\item Similar to earlier works, such as \citet{Pasquali2010}, we find that satellites are both more metal-rich ($< 0.1$~dex) and older ($< 2$~Gyr) than centrals of the same stellar mass, which has often been interpreted as an indication that the environment plays an important role in shaping both the chemical evolution and star formation histories of galaxies. However, after separating star-forming, green valley and passive galaxies, we find that the true environmental dependence is in fact much weaker, with star-forming, green valley and passive satellites being only marginally more metal-rich ($< 0.03$~dex) and older ($< 0.5$~Gyr) than star-forming, green valley and passive centrals of the same stellar mass.
\item The strong environmental effects (found by previous studies) when galaxies are not differentiated result from a selection effect brought about by the environmental dependence of the quenched fraction in galaxies, as both the stellar metallicities and stellar ages of passive galaxies are considerably larger than those of star-forming galaxies of the same stellar mass (with $\Delta \log Z_*\sim 0.15$ and $\Delta \mathrm{age}\sim$ 3~Gyr). Thus, we strongly advocate for the separation of the star-forming, green valley and passive populations of galaxies when the environmental dependence of galaxy properties are investigated, as otherwise both the true environmental dependence of these properties will be exaggerated and the mass dependence of any environmental effects will be misrepresented, especially when the properties are starkly different for star-forming and passive galaxies (e.g.\@ galaxy colours, $\mathrm{SFR}$s, galaxy sizes). 
\item We also investigate further environmental trends for both centrals and satellites separately, by dividing star-forming, green valley and passive galaxies into quartiles of halo mass, local overdensity and (for satellites only) projected distance from the central. We obtain the following results:
\begin{itemize}
\item We find no further environmental trends for star-forming galaxies, neither for centrals nor for satellites. 
\item In contrast, the stellar metallicities of green valley and passive satellites increase weakly ($<0.08$~dex and $<0.05$~dex, respectively) with increasing halo mass, increasing local overdensity and decreasing projected distance. 
\item Furthermore, we find that the stellar metallicities of central galaxies do not depend on the local overdensity. 
\item We also find a unique feature in the stellar mass--stellar metallicity relation for passive centrals, where galaxies in more massive haloes have larger stellar mass ($\sim$$0.1$~dex) at constant stellar metallicity. This effect is interpreted in terms of dry merging of passive central galaxies and/or progenitor bias.
\end{itemize} 
\item We use the method of \citet{Peng2015} and \citet{Trussler2020a} to investigate environmental quenching, using the stellar metallicity difference between star-forming and passive galaxies as a proxy for the prevalence of starvation as a quenching mechanism in galaxies. We obtain the following results:
\begin{itemize}
\item Following the lack of environmental trends in the stellar mass--stellar metallicity relations, we conclude that that the starvation responsible for quenching central galaxies is not driven by environmental phenomena, but primarily by mass-related phenomena. 
\item Additionally, we find that environmental starvation (`strangulation') only contributes moderately to the quenching of satellite galaxies, with the bulk of their starvation being due to mass-related effects. \end{itemize}
\item Finally, in order to put the previous findings on a more quantitative ground, we use the chemical evolution model of \citet{Trussler2020a} to derive constraints on environmental quenching. We find the following:
\begin{itemize}
\item We find that the durations of quenching $t_\mathrm{quench}$, which specify how long it takes for a galaxy to transition from star-forming to passive, are roughly independent of both mass and environment at 6--7~Gyr. The quenching durations for centrals and satellites are comparable at all stellar masses, and show no dependence on the local overdensity $\delta_5$.
\item Similarly, we find that the $e$-folding time-scales for quenching $\tau_\mathrm{q}$, which specify the time-scale over which the $\mathrm{SFR}$ decreases by a factor of $e$, are also roughly independent of both mass and environment at $\sim$1.2~Gyr. Again, the $e$-folding times for centrals and satellites are comparable at all stellar masses, and show no dependence on the local overdensity $\delta_5$.
\item The `effective' mass-loading factors $\lambda_\mathrm{eff}$, which describe the net gas expulsion rate ($\Lambda = \lambda_\mathrm{eff}\times\mathrm{SFR}$) during quenching, for central galaxies show no dependence on the local overdensity $\delta_5$.
\item In contrast, the `effective' mass-loading factors $\lambda_\mathrm{eff}$ for satellite galaxies (at all masses) decrease with increasing local overdensity $\delta_5$, with the quenching of satellite galaxies in high density environments being relatively more starvation-dominated than for satellite galaxies of the same stellar mass in low density environments (which are relatively more gas-stripping-/outflow-dominated). Furthermore, the strength of this environmental effect decreases with increasing stellar mass, from $3\sigma$ at $M_* = 10^{9.8}~\mathrm{M}_\odot$ to $< 0.5\sigma$ at $M_* = 10^{11}~\mathrm{M}_\odot$. 
\item Thus, our results indicate that it is in fact starvation, rather than gas stripping, that becomes progressively more important in quenching star formation in satellites in progressively denser environments. Furthermore, while environmental effects contribute moderately to the starvation and/or confinement of gas in low-mass satellite galaxies in dense environments, environmental effects only play a minor role in the evolution of massive satellites and satellites in low-density environments.
\end{itemize}
\end{enumerate}

\section*{Acknowledgements}

We thank the anonymous referee for their comments and suggestions which contributed to greatly improving this article. JT and RM acknowledge support from the ERC Advanced Grant 695671 `QUENCH'. RM acknowledges support by the Science and Technology Facilities Council (STFC). 
YP acknowledges the National Key R\&D Program of China, Grant 2016YFA0400702 and NSFC Grant No.\@ 11773001, 11721303, 11991052.
Funding for the SDSS and SDSS-II has been provided by the Alfred P. Sloan Foundation, the Participating Institutions, the National Science Foundation, the U.S. Department of Energy, the National Aeronautics and Space Administration, the Japanese Monbukagakusho, the Max Planck Society, and the Higher Education Funding Council for England. The SDSS Web Site is http://www.sdss.org/. 
The SDSS is managed by the Astrophysical Research Consortium for the Participating Institutions. The Participating Institutions are the American Museum of Natural History, Astrophysical Institute Potsdam, University of Basel, University of Cambridge, Case Western Reserve University, University of Chicago, Drexel University, Fermilab, the Institute for Advanced Study, the Japan Participation Group, Johns Hopkins University, the Joint Institute for Nuclear Astrophysics, the Kavli Institute for Particle Astrophysics and Cosmology, the Korean Scientist Group, the Chinese Academy of Sciences (LAMOST), Los Alamos National Laboratory, the Max-Planck-Institute for Astronomy (MPIA), the Max-Planck-Institute for Astrophysics (MPA), New Mexico State University, Ohio State University, University of Pittsburgh, University of Portsmouth, Princeton University, the United States Naval Observatory, and the University of Washington.

\section*{Data availability}

The SDSS galaxy parameter data used in this article are available at the MPA-JHU DR7 data release website: https://wwwmpa.mpa-garching.mpg.de/SDSS/DR7/. The SDSS group data used in this article can be downloaded from https://gax.sjtu.edu.cn/data/Group.html. The stellar population data underlying this article will be shared on reasonable request to the first author.




\bibliographystyle{mnras}
\bibliography{envDR7.bib,Supplementary.bib} 





\appendix

\section{Light-weighted stellar mass--stellar metallicity relation} \label{sec:mzr_lw}

\begin{figure*}
\centering
\centerline{\includegraphics[width=1\linewidth]{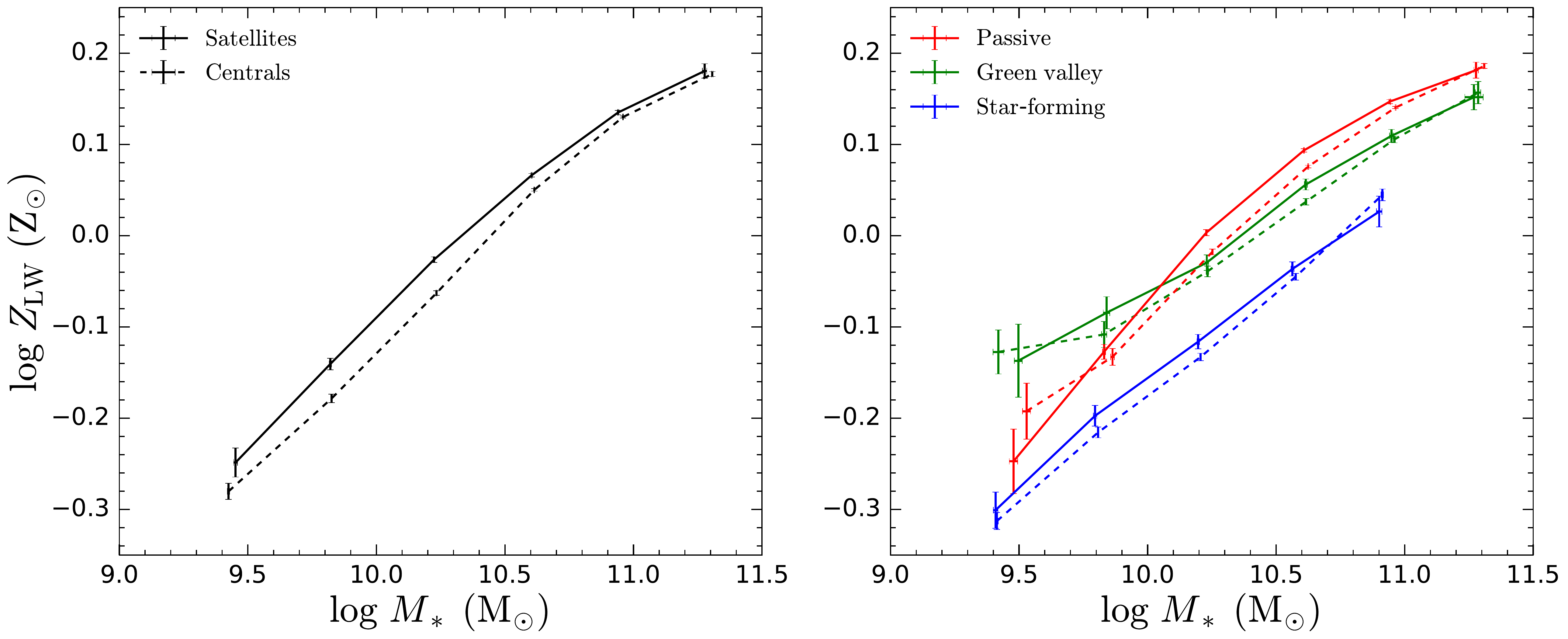}}
\caption{Similar to Fig.\@~\ref{fig:mzr_mw_cent_sat}, but now using light-weighted stellar metallicities from FIREFLY.}
\label{fig:mzr_lw_cent_sat}
\end{figure*}

\begin{figure*}
\centering
\centerline{\includegraphics[width=1\linewidth]{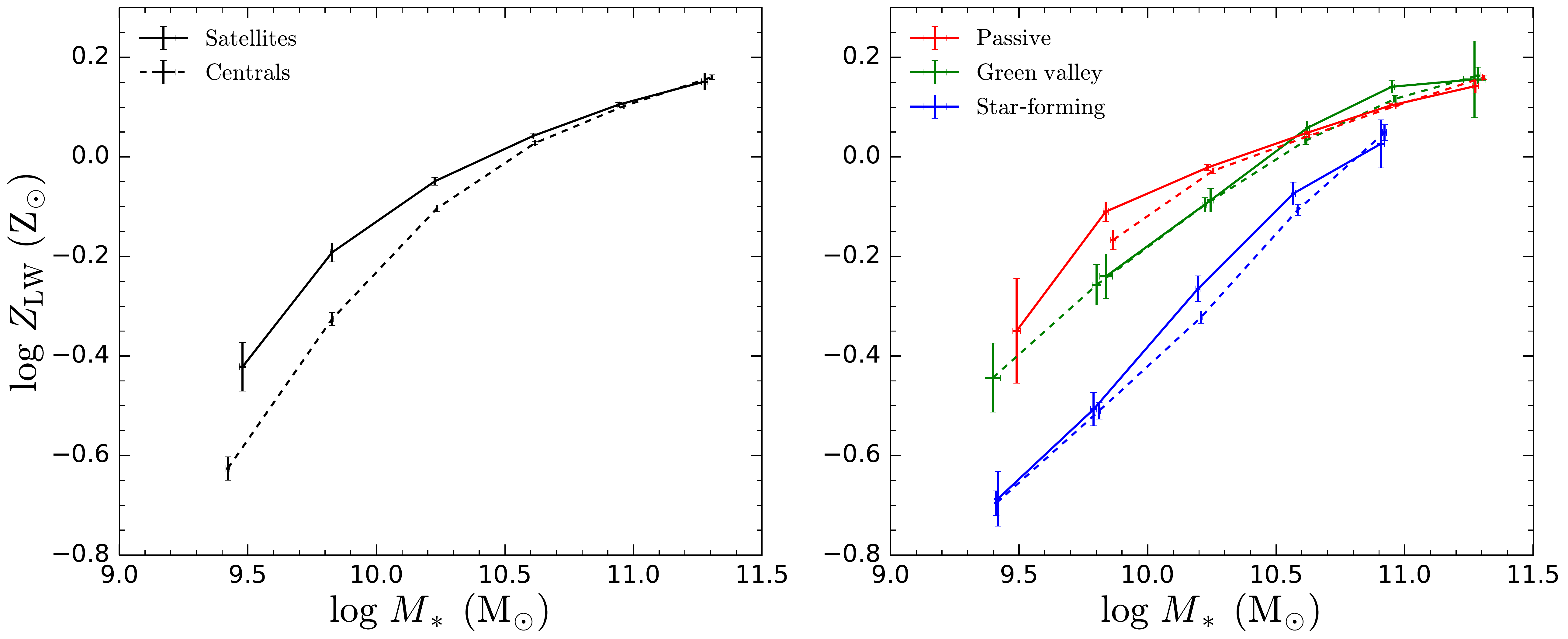}}
\caption{Similar to Fig.\@~\ref{fig:mzr_mw_cent_sat}, but now using light-weighted stellar metallicities from \citet{Gallazzi2005}.}
\label{fig:mzr_lw_g_cent_sat}
\end{figure*}

Our analysis in the main body of the text was centred on the mass-weighted stellar mass--stellar metallicity relation, using stellar metallicities that were obtained using the spectral fitting code {\footnotesize FIREFLY}. In order to investigate whether our results are affected by the weighting scheme or spectral fitting procedure adopted,  we study, in this section, the light-weighted stellar mass--stellar metallicity relation, using stellar metallicities that have been obtained using two different spectral fitting procedures. First, using a full spectral fit of the optical spectrum using {\footnotesize FIREFLY}. Second, using a simultaneous fit of five metallicity- and age-sensitive optical spectral absorption features \citep[for details see][]{Gallazzi2005}. We note that the latter set of stellar metallicities was used in the environment analysis of \citet{Pasquali2010}. 

We show the light-weighted stellar mass--stellar metallicity relation using the {\footnotesize FIREFLY} metallicities in Fig.\@~\ref{fig:mzr_lw_cent_sat}. Similar to what was seen for the mass-weighted stellar mass--stellar metallicity relation in Fig.\@~\ref{fig:mzr_mw_cent_sat}, satellites are typically more metal-rich than centrals of the same stellar mass. This result applies when studying the overall central and satellite population (left panel, in black) and also when studying the star-forming (right panel, blue), green valley (green) and passive (red) populations separately. We also find, as was seen for the mass-weighted stellar metallicities, that the metallicity difference between the overall population of centrals and satellites is larger than that for the star-forming, green valley and passive subpopulations. However, while the light-weighted central--satellite stellar metallicity differences (typically 0--0.02 dex) for star-forming, green valley, and passive galaxies are comparable to the mass-weighted differences (typically 0--0.03 dex), the light-weighted differences for the overall population are much smaller than the mass-weighted differences. This comes about because the stellar metallicity difference between star-forming, green valley and passive galaxies is smaller when using light-weighted metallicities over mass-weighted metallicities \citep[for a more in-depth discussion of this, see][]{Trussler2020a}, and so the quenched fraction effect only introduces a small, additional metallicity offset between the overall central and satellite populations for the light-weighted metallicities. 

Furthermore, we find that the metallicity difference between the overall population of centrals and satellites decreases with increasing stellar mass. In contrast, the metallicity difference for star-forming and green valley galaxies is roughly independent of stellar mass, and declines more weakly with mass for passive galaxies. These trends are all consistent with what was seen in our analysis of the mass-weighted stellar mass--stellar metallicity relation. Thus the choice of metallicity weighting used does not affect the qualitative aspects of our results. 

\begin{figure*}
\centering
\centerline{\includegraphics[width=1\linewidth]{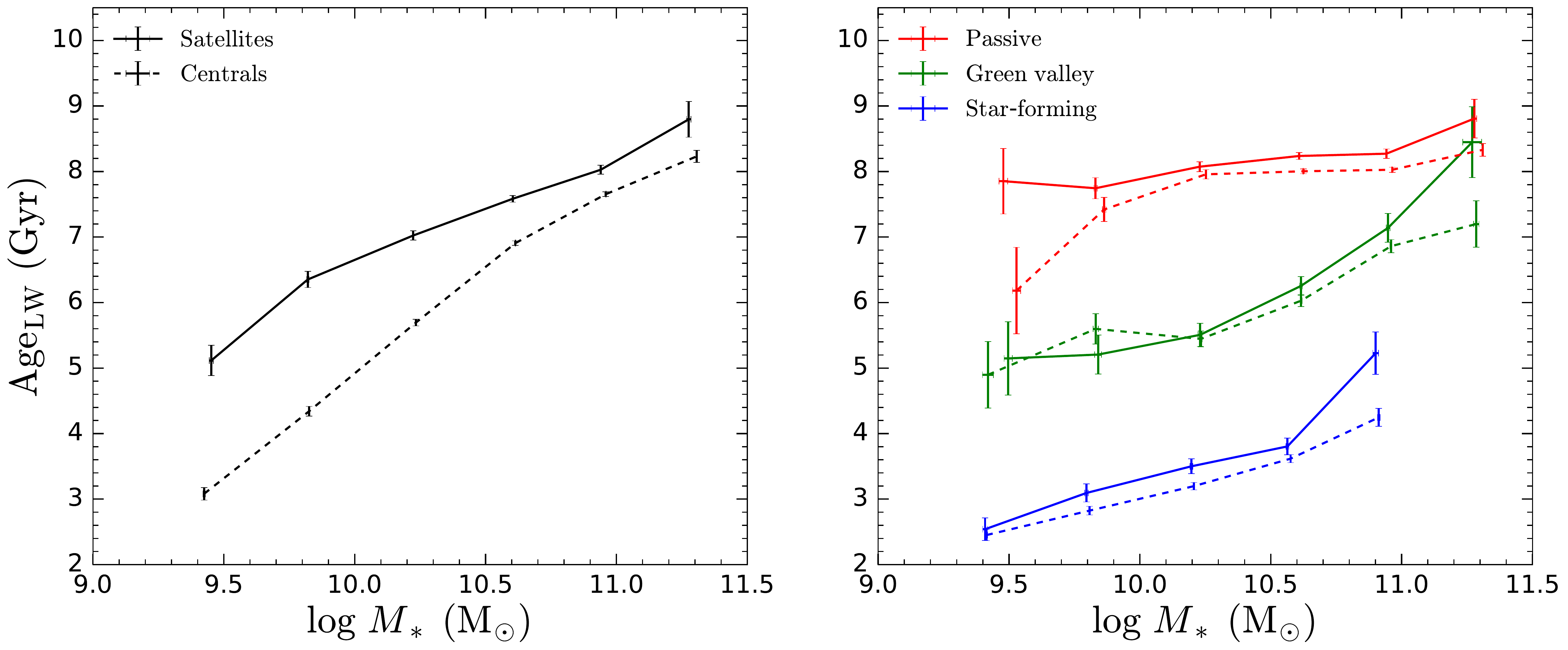}}
\caption{Similar to Fig.\@~\ref{fig:mar_mw_cent_sat}, but now using light-weighted stellar ages from FIREFLY.}
\label{fig:mar_lw_cent_sat}
\end{figure*}

\begin{figure*}
\centering
\centerline{\includegraphics[width=1\linewidth]{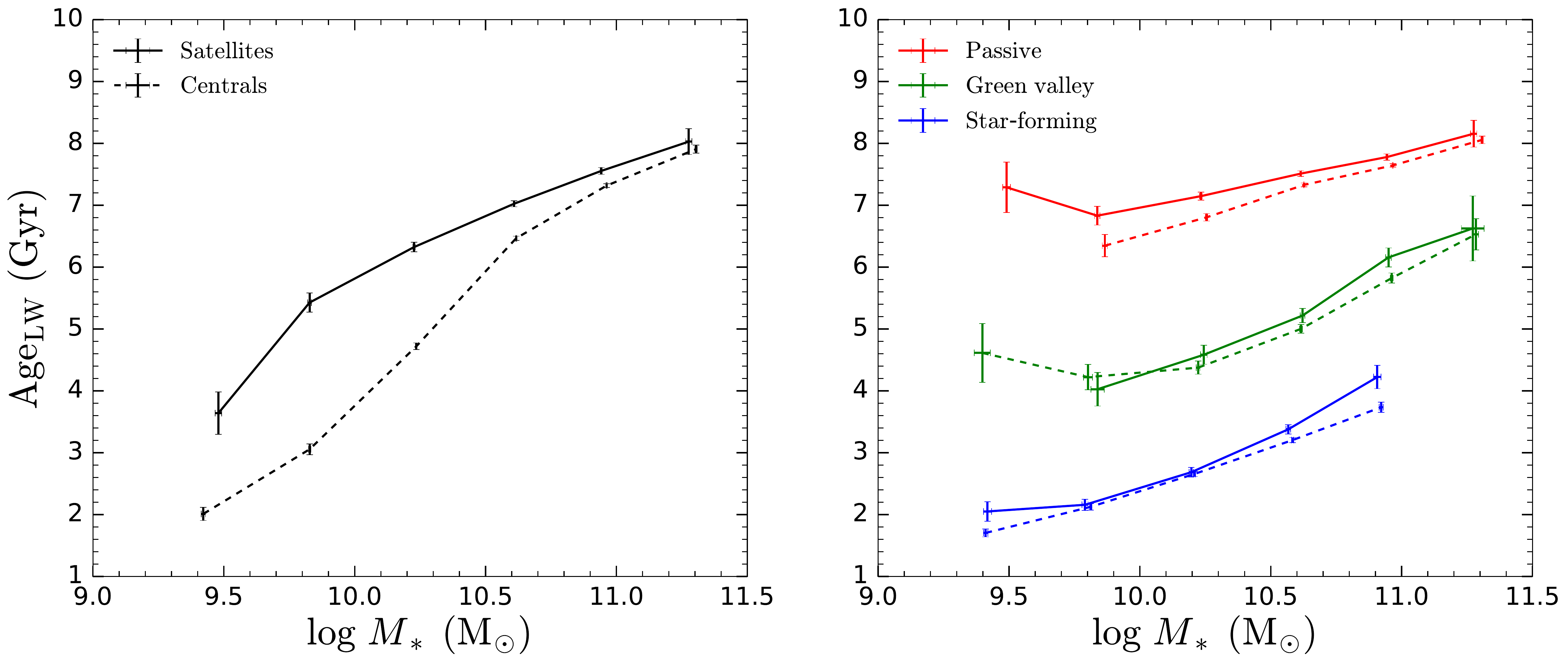}}
\caption{Similar to Fig.\@~\ref{fig:mar_mw_cent_sat}, but now using light-weighted stellar ages from \citet{Gallazzi2005}.}
\label{fig:mar_lw_g_cent_sat}
\end{figure*}

The light-weighted stellar mass--stellar metallicity relation using the \citet{Gallazzi2005} metallicities is shown in Fig.\@~\ref{fig:mzr_lw_g_cent_sat}. Qualitatively, we find the same trends as with the {\footnotesize FIREFLY} metallicities. Thus, the choice of spectral fitting procedure used also does not appear to affect the qualitative aspects of our results. However, we do find that the quantitative aspects of our results are changed, especially for the stellar metallicity difference between the overall population of centrals and satellites, which we find to be much larger when using the \citet{Gallazzi2005} metallicities. This comes about because the normalisations of the stellar mass--stellar metallicity relations based off of {\footnotesize FIREFLY} and \citet{Gallazzi2005} are different \citep[which is discussed in detail in][]{Trussler2020a}. Since the stellar metallicity difference between star-forming, green valley and passive galaxies is much larger when using the \citet{Gallazzi2005} metallicities, the quenched fraction effect introduces a much larger, additional metallicity offset between the overall central and satellite populations, which is why we see such a large gap between centrals and satellites in the left panel of Fig.\@~\ref{fig:mzr_lw_g_cent_sat}.

\section{Light-weighted stellar mass--stellar age relation} \label{sec:mar_lw}

In this section we study the light-weighted stellar mass--stellar age relation, using stellar ages obtained using {\footnotesize FIREFLY} and from \citet{Gallazzi2005}, to test whether our results are affected by the weighting scheme or spectral fitting procedure used.

We show the light-weighted stellar mass--stellar age relation using the {\footnotesize FIREFLY} ages in Fig.\@~\ref{fig:mar_lw_cent_sat}. Qualitatively, we find similar trends to what was found with the mass-weighted stellar mass--stellar age relation. Satellites tend to be older than centrals of the same stellar mass, both for the overall population (shown in the left panel) and for star-forming (blue, shown in the right panel), green valley (green) and passive (red) galaxies, the quenched fraction effect exaggerates the true (shown on the right) age difference between centrals and satellites, and the quenched fraction effect also misrepresents the stellar-mass dependence of the true stellar age difference between centrals and satellites. Since the gap in stellar age between star-forming, green valley and passive galaxies is larger when using light-weighted over mass-weighted ages, the quenched fraction effect is stronger for light-weighted ages and so the age difference between the overall central and satellite populations are larger than in the mass-weighted case.

The light-weighted stellar mass--stellar age relation using the \citet{Gallazzi2005} ages is shown in Fig.\@~\ref{fig:mar_lw_g_cent_sat}. The trends obtained are qualitatively consistent with what was found using the light-weighted {\footnotesize FIREFLY} ages, so the choice of spectral fitting procedure adopted does not appear to affect the qualitative aspects of our results.

\section{Comparison with gas-phase metallicity studies} \label{sec:discussion}

In this section we discuss the environmental trends that we have found for stellar metallicities, focussing on the comparison against environmental trends that previous studies have found for \textit{gas-phase} metallicities. It should be noted that while gas-phase metallicities have only been measured for star-forming galaxies, stellar metallicities can be measured for both star-forming and passive galaxies. 

The environmental dependence of the stellar populations in galaxies has previously been investigated by \citet{Pasquali2010}, who studied the overall populations of centrals and satellites and did not differentiate between star-forming, green valley and passive galaxies in their analysis. They found that the stellar metallicities of galaxies are strongly dependent on environment, with satellites being substantially more metal-rich ($0$--$0.2$~dex, decreasing with stellar mass) than centrals of the same stellar mass. In contrast, \citet{Pasquali2012} studied the gas-phase metallicities in star-forming galaxies, and found that the environmental dependence was much weaker, with satellites only being marginally more metal-rich ($0$-$0.05$~dex, decreasing with stellar mass) than centrals of the same stellar mass. Furthermore, after restricting their stellar metallicity analysis to a sample of galaxies for which gas-phase metallicities are available (i.e.\@ only including star-forming galaxies and excluding passive galaxies), \citet{Pasquali2012} found that the stellar metallicity difference between centrals and satellites becomes much smaller ($\sim$$0$).  Thus the large stellar metallicity differences between centrals and satellites found in \citet{Pasquali2010} were primarily driven by the environmental dependence of the quenched fraction of galaxies. As shown in this work \citep[and implicitly by the findings of][]{Pasquali2012}, the true stellar metallicity difference between centrals and satellites is in fact much smaller (see Fig.\@~\ref{fig:mzr_mw_cent_sat}).

In addition to \citet{Pasquali2012}, numerous other works have also compared the gas-phase metallicities of central and satellite galaxies, finding that satellites are typically more metal-rich than centrals of the same stellar mass \citep[see e.g.\@][]{Ellison2009a, Peng2014a}. Interestingly, although this is qualitatively similar to our results for star-forming galaxies (as well as green valley and passive galaxies), we note that the \textit{stellar} metallicity differences we find between star-forming centrals and satellites ($0.01$--$0.02$~dex) are smaller than the corresponding \textit{gas-phase} metallicity differences ($\sim$$0.05$~dex). This quantitative disagreement can come about for two reasons. First, systematic effects associated with the choice of metallicity calibration and the different methods for measuring gas-phase and stellar metallicities. Second, while gas-phase metallicities provide an instantaneous measure of chemical enrichment as they probe the current conditions in the ISM, stellar metallicities instead provide a cumulative measure of chemical enrichment that is averaged over longer timescales (especially for the mass-weighted metallicities we study in our analysis). As a result, stellar metallicities will react less promptly and therefore be less sensitive to recent or ongoing environmentally-driven processes that cause metal enrichment, due to the averaging effect over the entire stellar population. 

The dependence of gas-phase metallicities on further environmental parameters has also been investigated, with studies finding that the gas-phase metallicities of star-forming satellites at a fixed stellar mass tends to increase (by $<0.05$~dex) with increasing halo mass \citep[e.g.\@][]{Pasquali2012}, increasing local overdensity \citep[e.g.\@][]{Mouhcine2007, Cooper2008, Ellison2009a, Peng2014a, Wu2017a} and decreasing projected distance from the central galaxy \citep[for massive clusters, e.g.\@][]{Petropoulou2012, Maier2016, Maier2019}. In contrast, we find that the stellar metallicities of star-forming satellites do not depend on environment. This discrepancy between the trends for gas-phase and stellar metallicities is likely to be due to the fact that stellar metallicities react less promptly and are therefore less sensitive to recent metal enrichment (which is likely being driven by environmental processes). However, if the enrichment (process) persists on longer timescales, especially during the quenching process, then this environmental dependence should become observable in the stellar metallicities as well, especially for passive satellites and potentially also for green valley satellites. Indeed, we find that the stellar metallicities of passive satellites (and green valley satellites, although with less significance) increase (by $<0.05$~dex and $<0.08$~dex for passive satellites and green valley satellites, respectively) with increasing halo mass, increasing local overdensity and decreasing projected distance, which is likely reflecting the aforementioned environmental trends for the gas-phase metallicities of star-forming satellites. 

Furthermore, both \citet{Peng2014a} and \citet{Lian2019a} find that the environmental dependence (in terms of the local overdensity) of the gas-phase metallicities of star-forming satellites decreases with increasing stellar mass. Again, while we do not see this trend in the stellar metallicities of star-forming satellites, it is readily apparent for passive satellites. Indeed, we find that, for both the local overdensity and projected distance, the environmental dependence of the stellar metallicities of passive satellites decreases with increasing stellar mass.

Finally, similar to studies of the gas-phase metallicities of central galaxies \citep[e.g.\@][]{Peng2014a}, we find that the stellar metallicities of star-forming centrals (and also green valley and passive centrals) do not depend on local overdensity.

\bsp	
\label{lastpage}
\end{document}